# ALEXIS: Recreating the X-Ray Emission from the Full Solar Disk as a Linear Combination of Discrete Regions in the Extreme Ultraviolet and Soft X-Rays

Jorge R. Padial Doble[1] and Kelly Holley-Bockelmann[1,2]

[1] Vanderbilt University, Physics and Astronomy Department, Nashville, TN 37209, USA; alexissolarflarecatalog@gmail.com
[2] Department of Physics, Fisk University, 1000 17th Ave. N, Nashville, TN 37208, USA

## Abstract

Despite a wealth of multiwavelength, spatially resolved, time-domain solar activity data, an accurate and complete temporospatial solar flare census is unavailable, which impedes our understanding of the physics of flare production. We present an Automatically Labeled EUV and X-Ray Incident SolarFlares (ALEXIS) pipeline, designed to decompose the X-ray flux of the full solar disk into a minimum set of discrete regions on the solar surface. ALEXIS returns an average rms error between the X-Ray Sensor time series and the discrete EUV signals of $0.066 \pm 0.036$ for a randomly selected test bed sample of 1000 hr-long data segments from 2010 May to 2020 March. Flare emission that requires multiple regions was found to be synchronous (flares occurring at the same time), sympathetic (flares separated by minutes), or needed to capture the background emission before and/or after the main flare. ALEXIS uses the original full resolution and cadence of both the Atmospheric Imaging Assembly instrument and the GOES13–15 Solar X-Ray Imager. Comparison of ALEXIS's catalog with those produced by the SWPC and SolarSoft show that these canonical databases need revisiting for 62% and 15% of the subsample, respectively. Additionally, we increased the number of flares reported by the SWPC and SolarSoft by 15%. Our pipeline misses 6.7% of the 1057 flare subsample and returns 5% of false positives from 1211 flares reported by ALEXIS. The ALEXIS catalog returns flare peak times, coordinates, the corrected scaled X-ray magnitude, and the associated NOAA active region with a HARP identifier number independently from any external data products.

*Unified Astronomy Thesaurus concepts:* Solar flares (1496); Solar extreme ultraviolet emission (1493); Solar x-ray emission (1536)

## 1. Introduction

Solar flares are characterized by a rapid increase in radiative flux followed by a slow decline period that occurs on the outer surfaces of the Sun across the electromagnetic spectrum. These flares are localized mostly within large, complex, rapidly evolving, and magnetically unstable areas called active regions (ARs). Large solar flares and events associated with them, like coronal mass ejections (CMEs), are rare but highly destructive phenomena that can impact social, economic, and digital infrastructure on Earth. Eruptive events are dangerous because they can propel charged particles throughout the solar system, eventually interacting with planetary magnetic fields. On Earth, strong storms such as those observed by Carrington in 1859, the event in 1989 that interrupted Quebec's grid, or the event that missed Earth in 2012 can cause disruptions of communications, electric grids, harm to astronauts, and affect the thermographic drag experienced by commercial and government-owned satellites (P. M. Mehta & R. Linares 2018; D. H. Boteler 2019; H. S. Hudson 2021; T.-W. Fang et al. 2022; R. J. Licata & P. M. Mehta 2022). Forecasting when, where, and how strong a solar storm can occur can save lives and trillions of dollars in infrastructure damage, and avoid international misunderstandings that can lead to nuclear escalation

(H. S. I. D. Bäumen et al. 2014; D. J. Knipp et al. 2016; J. P. Eastwood et al. 2017).

Several data products exist to localize, characterize, and catalog flaring events and ARs. All flare catalogs have one common starting point: an X-ray class or label. This logarithmic scale encodes the strength of the flare by assigning an alphanumerical classification (A-B-C-M-X and 1.0–9.9). Combinations of a letter and a number map to the X-ray flux of the full solar disk at the time the flare peaked. The largest events (X- and M-class flares) are rare while smaller events (<C class) are more common. Some other characteristics tracked by these catalogs include the flare start, stop, and peak times, which are needed to describe the solar flare class, coordinates, and the AR it belongs too.

One of the most relied upon data products is a solar flare catalog compiled by the Space Weather Prediction Center (SWPC), which describes the location and X-ray magnitude of solar flares. The SWPC catalog defines solar flares using the X-ray flux integrating over the full solar disk and cannot spatially resolve explosive events. The Solar X-ray Imager (SXI) and other ground-based telescopes are then used to locate these eruptive events. Larger flares are easier to identify, while the opposite is true for fainter flares. On the other hand, the Lockheed Martin Solar Astrophysics Lab (SolarSoft) produces another catalog that depends on the event list produced by the SWPC. SolarSoft searches the Atmospheric Imaging Assembly (AIA) bands and fills the gap in spatial information, but their methods are unpublished. The Stanford group produces a third canonical data product that describes the dynamic magnetic properties of ARs (M. G. Bobra et al. 2014). Their tracked ARs when combined with magnetic field metadata produce the Space Weather HMI Active Region Patches (SHARPs). The Helioseismic and Magnetic Imager (HMI) reports a time series of metadata and tracks how they evolve as the AR progresses across the solar disk over a period of ≈14 days. Nevertheless, to label an AR and its associated SHARP as flaring or nonflaring, they still depend on the localization and characterization of events by the SWPC and SolarSoft. These catalogs and data products have been useful and have guided decades of heliophysics research that includes, but is not limited to, machine learning (ML) approaches applied to the prediction of solar flares, but they still have limitations.

There have been many attempts to use different ML tasks applied to heliophysics. These attempts can be grouped into predictive models before and after the advent of the Solar Dynamics Observatory's (SDO) HMI instrument. In general, predictive models using HMI data try to predict parameters such as the flare index (H. Zhang et al. 2022) or the center coordinates and the solar radius using a convolutional neural network (G. Zhu et al. 2020). Other applications of ML have been applied to predict when and how the strongest solar eruptive events might occur. These classification tasks use data from multiple space-based telescopes and all have the same baseline assumption: scientists know when, where, and how strong a solar flare is. Scientists compile the SWPC flare catalog and pair the metadata with the SHARPs to develop models that range from simple binary classifiers to uninterpretable deep-learning models. All attempts, regardless of how much data they are given, yield similar mean evaluation metrics such as the true skill score or the area under the precision–recall curves (D. S. Bloomfield et al. 2012; X. Huang et al. 2013; M. G. Bobra & S. Couvidat 2015; T. Muranushi et al. 2015; C. Liu et al. 2017; N. Nishizuka et al. 2017; F. Benvenuto et al. 2018; K. Florios et al. 2018; X. Huang et al. 2018; E. Jonas et al. 2018; E. Park et al. 2018; Y. Chen et al. 2019; H. Liu et al. 2019; J. Wang et al. 2019; Y. Zheng et al. 2019; S. Hazra et al. 2020; Z. Jiao et al. 2020; X. Li et al. 2020; X. Wang et al. 2020; Y. Abduallah et al. 2021; N. Nishizuka et al. 2021; Z. Sun et al. 2022; B. Panos et al. 2023). The lack of increased performance with model complexity or the ingestion of larger training sets can be attributed to two main reasons.

First, we may have reached the pinnacle of ML approaches with the current scale/resolution of data available; we need new sets of physical descriptors. Second, the data that are being used are incomplete or incorrectly labeled. The location of a solar flare is important for prediction tasks because it allows us to associate flares to the magnetic field regions that produced them, which in turn determines the classification labels fed into ML models. To test the second hypothesis, we developed the Automatically Labeled EUV and X-Ray Incident SolarFlare (ALEXIS) pipeline. We present a data-driven approach that can create a solar flare catalog with specific coordinates. This catalog can better characterize the small (C-class) events in order to increase the amount of data that can be used in predictive models.

To achieve this task, ALEXIS uses data combined from the best space-based telescopes designed for heliophysics: the SDO and the Geostationary Operational Environmental Satellite (GOES) covering the extreme-ultraviolet and X-ray regions of the electromagnetic spectrum, respectively. Three main instruments are used: (1) the highest resolution and cadence available from the state-of-the-art AIA; (2) the SXI; (3) and the X-Ray Sensor (XRS). Solar flare locations are learned by ALEXIS by recreating the X-ray time series from the XRS with a weighted linear combination of discrete regions as observed by the multipixel AIA and SXI instruments.

The objective of this pipeline is to create a publicly available data product that increases the validity and veracity of science applied to solar flares, atmospheric heliophysics, and the Earth–Sun system. The primary purpose of this paper is to show a proof of method that independently verifies solar flares detected and cataloged by the SWPC and SolarSoft. ALEXIS uses contemporary computer vision, optimization, and scaled infrastructure applied to the full cadence and resolution of AIA and SXI images. The first version of the ALEXIS catalog will include 1000 1 hr time ranges where C-class solar flares have been identified by the SWPC, SolarSoft, and ALEXIS as compiled randomly from 2010 May to 2020 March.

In a second paper, ALEXIS will provide a downloadable database with the location, magnitude, and associated AR information for all flares NOAA has detected and failed to categorize correctly. The second paper will include the >8300 events parsing through hundreds of terabytes of image data between 2010 May and 2020 March (J. R. Padial-Doble 2025, in preparation). Preliminary results indicate that further development of ALEXIS beyond its current stage has the capacity to increase the total known solar flare catalog by ≈90%. Doing so would require computer resources that may only be accessible via the infrastructure provided by Argonne or Oak Ridge Leadership Computing Facilities.

This paper is organized in the following way. In Section 2, we describe the telescopes and instruments used. Section 3 describes the motivation and data used in this proof-of-concept methods paper. Section 4 describes how ALEXIS learns the discrete locations of solar flares and establishes their metadata. Section 5 shows the results of the pipeline and compares them to canonical databases. Section 6 discusses the implications of ALEXIS's catalog and poses questions that it might be able to answer. Section 7 summarizes the contributions. In the Appendices, you will find the data-cleaning procedures, visual examples of the flares in the catalog, a description of the databases produced, and supplemental figures for interpretability.

## 2. Telescopes and Instruments

### 2.1. GOES13–15

The GOES is a satellite that has had several iterations since 1975, which are assigned a numerical identification (GOES1–18). The GOES is operated by the National Oceanic and Atmospheric Administration (NOAA), where its primary function is to gather measurements that help in meteorological forecasting for the Americas, parts of the Pacific, and the Alaskan basin. Another reason the GOES has been a continuous mission is because it aids in alerting the government and research agencies about hazardous space weather conditions such as high-energy particles or changes in the near-Earth magnetic field. There are always two GOES telescopes in operation, a primary satellite and a secondary satellite, which are located at a height of 36,800 km and separated by 60°.

On the Sun-facing part of the GOES telescope, there are several instruments designed for research in heliophysics,

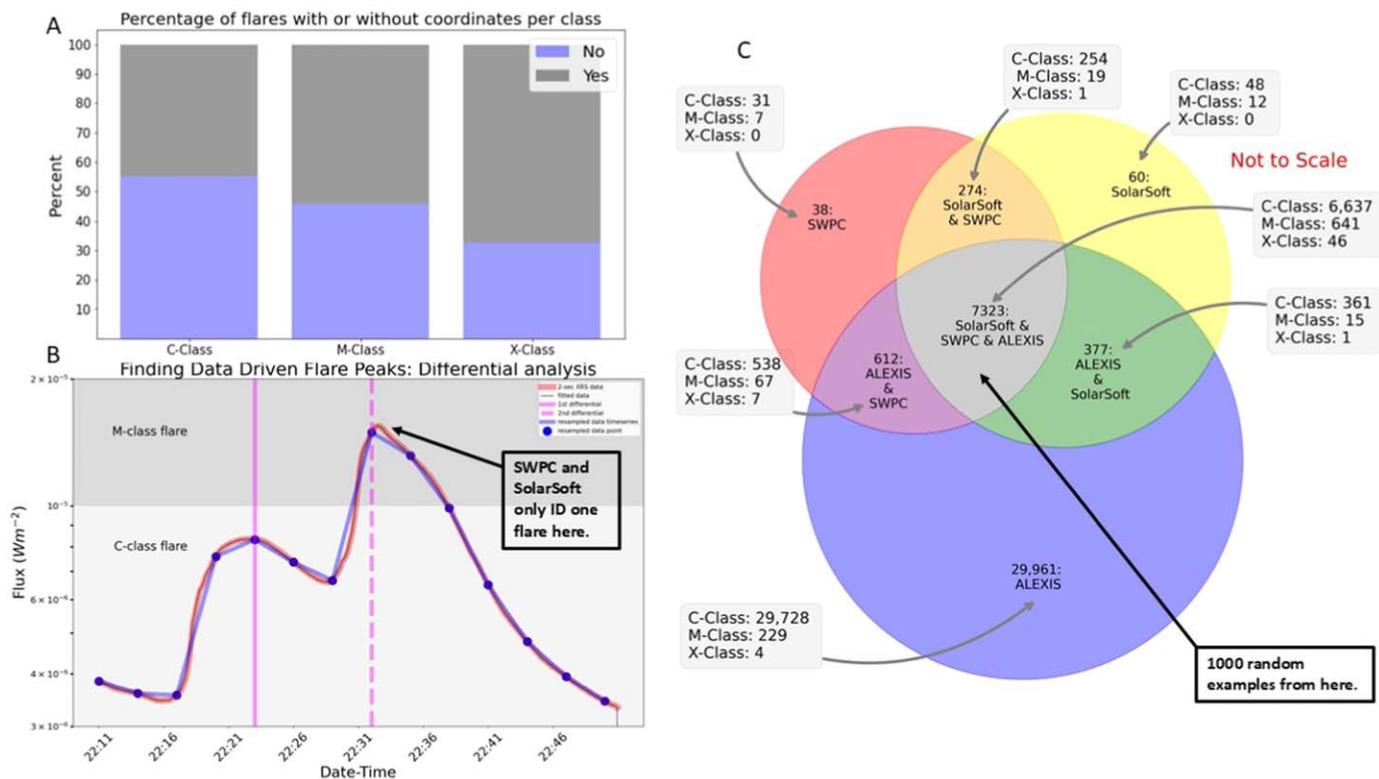

**Figure 1.** (A) Bar plot of GOES flares with or without locations from 2010 May to 2020 May: flares compiled from the Heliophysics Event Knowledgebase reported by GOES. They report if they have location metadata (purple) and if they do not have location metadata (gray). More than 50% of the C-class flares and above have no location metadata. (B) Differential analysis of the XRS signal: approximately 1 hr of data on 2014 January 7 at 22:31:30 UTC searching for a C7.4-class flare. Note that the XRS scale-corrected 2 s flux (maroon) will now push the magnitude of this flare into the M-class category; spline-fitted data (black); resampled data points at 3 minute cadence (blue scatter); time series of resampled points (blue line); magenta vertical lines show the local maxima. Note, SWPC and SolarSoft report one flare for this time range. ALEXIS returns two possible flares (the solid magenta vertical line is the first local maxima chronologically and the magenta dotted vertical line is the second local maxima). Upon completion of the pipeline, there are two flares that happened in this signal: the first is a C8.35 flare at 22:23:11 UTC with Helioprojective Cartesian (HPC) [−80″, −131″] associated with HARP 3563 close to the solar center and another M1.55 flare at 22:33:07 UTC with coordinates HPC [951″, 208″] close to the solar border and HARP 3587 (see Figure 5). (C) Distribution of all the local maxima found by differential analysis: SWPC (red), SolarSoft (yellow), and ALEXIS (purple). The differential technique recovers all but 372 events (4.3%). All of the missed events are recuperated when using the non-science-quality XRS data. ALEXIS identifies about 4 times the number of flares reported by other teams. This paper reports on a random sample of 1000 events from the center of the Venn diagram.

atmospheric sciences, and space weather. The Space Environment Monitor, one of the telescope's subsystems, has an X-ray imager (SXI) along with other high-energy, magnetic, and extreme-ultraviolet detectors dedicated to monitoring the Sun and the near-Earth geosynchronous environment. Together with the XRS, these data allow the team at the SWPC to "receive, monitor, and interpret" solar phenomena like solar flares or ARs. For this analysis, we will use data from GOES13 (GOES-N), GOES14 (GOES-O), and GOES15 (GOES-P). One or more of these instruments have been used interchangeably as primary and secondary telescopes and have taken mostly continuous operational data from September 2009 to March 2020.

### 2.1.1. XRS

Mounted on a gimbal of the X-Ray Positioner of the GOES solar yoke, the spacecraft will always point so that the XRS always faces the Sun. With a nominal field of view of 29.′1, the XRS measures the integrated flux of the full solar disk every 2.048 s in two spectral bands. These broadband channels, called XRS-A (0.5–4 Å) and XRS-B (1–8 Å), are commonly referred to as XRS long and XRS short wavelengths, respectively. These instruments are gas-filled iron chambers with ion cell detectors where a combination of filters are used to primarily detect and assign the X-ray label to flares (H. A. Garcia 1994; P. L. Bornmann et al. 1996). The XRS is a single-pixel detector that is unable to determine where the incoming X-rays originate on the solar surface but can point to the center of the Sun within ±2′. The spectral response of both XRS detectors are orders of magnitude more sensitive to flaring events than to the quiet Sun, as calculated with CHIANTI spectra. The saturation of the XRS detector occurs and is expected to be at values of $>10^{-3}$ W m$^{-2}$. XRS science-quality data are defined as the irradiances that have been preprocessed by the NOAA National Centers for Environmental Information (NCEI). Raw data are preprocessed to contain corrections for the gain, background electronic offsets, SWPC scaling factors, time stamps, conversion factors, and temperature effects. Please refer to the GOES13–15 read-me file[3] for more information. Note that ALEXIS only uses good science-quality XRS data; this is important when looking at Figure 1(C).

---

[3] https://www.ngdc.noaa.gov/stp/satellite/goes/doc/GOES_XRS_readme.pdf

### 2.1.2. SXI

The SXI was developed by Lockheed Martin with the objective of aiding the US Air Force and NOAA in creating solar forecasts and informing about conditions that may affect navigation equipment, satellite integrity, or astronaut security. The SXI is an imager part of the Solar Environment Monitor set of instruments whose objective is to locate coronal holes and solar flares while also monitoring coronal mass ejections and ARs. It takes 540 × 512 pixel images at a resolution of 5″ per pixel in a wavelength band that covers the soft X-rays to the extreme ultraviolet (6–60 Å or 0.2–2.1 keV). A list of predetermined exposure times helps the SXI monitor for different types of solar activity. The images that ALEXIS uses are those that have metadata that identify exposure times that are optimized for the study of solar flares (typical exposure time <10 ms). For the temporal continuity of these images, we only use the thin-poly (PTHN; poly/Al/Ti), thick-poly (PTHK; poly/Ai/Ti), and tin-mesh (TM; Poly/Sn) filters. These cover the wavelength ranges of 6–60 Å, 6–50 Å, and 6–70 Å, respectively.

To identify the center of the Sun, the SXI depends on the High Accuracy Sun Sensor (HASS) system. The HASS reports a nominal pointing knowledge of ±10″, but errors of up to 30″ have been reported. The errors in pointing have rendered this data set practically unusable. ALEXIS's cleaning technique finds that, for solar flare images from the SXI, the actual pointing error to the center of the Sun is on average 64″ (see Appendix A). Our preprocessing technique has facilitated the use of the SXI with popular Python software packages such as Astropy and Sunpy; metadata for SXI fits files are corrected by ALEXIS.

## 2.2. SDO

Launched in February 2010 as part of NASA's Living With a Star Program, the SDO (W. D. Pesnell et al. 2012) aims to address heliophysics science that requires high temporal and spatial resolution. Its main objective is to provide data to understand the mechanisms behind solar activity in order to answer questions related to how the Sun generates, stores, and releases magnetic energy. Other objectives include the study of the rapid evolution of the solar atmosphere and the origins of space weather. Over the past 14 yr, the SDO has continuously observed the Sun and has produced more than 25 PB of data, which are stored by the Joint Science Operations Center (JSOC) hosted at Stanford University. These data have been used to probe the solar structure, dynamics, and processes that enhance and produce solar activity, such as flares, coronal mass ejections, solar prominences, and coronal loops. There are three main instruments on board the SDO: the HMI, the AIA, and the Extreme Ultraviolet Variability Experiment. Note that specific solar events can be queried using the Heliophysics Event Knowledgebase (HEK; N. Hurlburt et al. 2012), which is hosted by SolarSoft.

### 2.2.1. HMI

The HMI is an instrument on board the SDO, created to study the interior and the magnetic variability of the Sun (P. H. Scherrer et al. 2012; J. Schou et al. 2012). The HMI returns 4096 × 4096 pixel images at a resolution of 0.″6 per pixel at 45, 90, or 135 s cadences. This instrument measures the vector magnetic field at the photosphere using the Doppler shift, intensity, and polarization of the 6173 Å Fe absorption line. Science products from the HMI allow for the study of the convective movement of the solar interior that generates a magnetic field capable of twisting, shearing, and moving toward the solar surface. An automatic pipeline has been created to spatially track magnetic structures as they cross the solar disk using line-of-sight magnetograms at 720 s cadence, called HMI Active Region Patches (HARPs; J. T. Hoeksema et al. 2014). The automatically detected and tracked HARP magnetic structures can vary in size and can engulf one or more NOAA AR members. Basic characteristics of the HARP are also returned, such as the total area of the AR and the net line-of-sight magnetic flux.

A more sophisticated data product that combines HARPs with calculated metadata that summarizes physical values was presented by M. G. Bobra et al. (2014). The SHARPs provide a more complete data product that combines HARPs with calculated metadata summarizing physical values reported every 12 minutes. These physical values have been associated with imminent flaring activity such as magnetic helicity, magnetic shear angle, or other free energy proxies (K. D. Leka & G. Barnes 2007). Currently, no single or combination of parameters has been found to be sufficient to predict whether a solar AR has more probability to produce a flare.

Any researcher interested in using the SHARPs for solar flare prediction must map from the metadata in an event list to the HARP that produced it. In the field of solar flare prediction, flare-finding teams report to the HEK a list of flares that provide temporal and spatial metadata. Spatial metadata include specific pixel coordinates, a boundary box, and the NOAA AR associated with the flare entry. In practice, researchers map from the event's associated NOAA AR to a HARP that has been associated with that NOAA AR number. It should be noted that multiple NOAA ARs can map to a single HARP and that the same is true for multiple HARPs mapping back to a single NOAA AR (M. G. Bobra & S. Couvidat 2015).

### 2.2.2. AIA

The AIA is one of the key instruments on board the SDO whose capabilities are ideal for the study of small structures with high variability. The purpose of the AIA is to study the dynamics associated with the evolution of solar flares, coronal mass ejections, and coronal loops. The AIA also increased scientific studies on the impact of solar activity on space weather and its effects on Earth's magnetosphere and ionosphere. The AIA (J. R. Lemen et al. 2012) returns 16 megapixel (4096 × 4096) images in extreme-ultraviolet and optical of the full Earth-facing solar disk. The fast-cadence AIA filters allow us to study the photosphere, chromosphere, transition region, and the different layers of the corona. These images span 10 wavelengths, are returned at different cadences (12, 24, and 3600 s), each filter has degraded independently of one another over time, and the wavelength response function of some filters overlap (J. Schou et al. 2012). For more information on the high-cadence band passes used in this paper, refer to Table 1.

**Table 1**
Reference for Details Regarding the AIA Fast-cadence Filters Corona: 94, 171, 193 (Iron 12), 211, 335; Transition Region/Chromosphere: 304, 171; Flares: 131, 193, 94; ARs: 211, 335

| Wavelength | Temperature | Abundance | Comments |
|---|---|---|---|
| 94 Å | 6,000,000 K | Fe XVIII<br>Fe XXVIII | Corona during flare |
| 131 Å | 10,000,000 K | Fe XX<br>Fe XXIII | Flaring material<br>Flaring regions |
| 171 Å | 600,000 K | Fe IX | Quiet corona<br>Coronal loops<br>Upper transition region |
| 193 Å | 1,000,000 K<br>20,000 K | Fe IX<br>Fe XXIV | Slightly hotter corona<br>Solar flare material |
| 211 Å | 2,000,000 K | Fe XIV | Hot ARs at corona |
| 304 Å | 50,000 K | He II | Chromosphere<br>Transition region |
| 335 Å | 2,500,000 K | Fe XVI | Hot ARs at corona |

Note that the readout characteristics of the CCD detector will cause the automatic exposure correction (AEC) algorithm to change the exposure of the imagers until the image is no longer saturated. The saturation point of the AIA sensor occurs at pixel values of $2^{14} = 16,384$, given that it is a 14-bit detector. The AEC changes the exposure time of each subsequent image to values that span 0.2–3 s intermittently. In other words, if an image is saturated, the AEC will lower the exposure for the subsequent image. After the short-exposure image, it tries a longer exposure and asks if the image is still saturated. If the image is saturated, it reverts to a short exposure for the next image. Next, a long exposure. This interchange of exposures continues until the image is no longer saturated.

The average file size for an AIA image is 10 MB. Once the image is opened and cleaned, the resulting files are 10 times the size. The amount of filters the AIA allows for and the cadence of the data stream make using the full data product difficult and require large data storage resources. Some data products exist that condense AIA data into more manageable data cubes. For example, K. Dissauer et al. ([2023](#)) and R. Galvez et al. ([2019](#)) condense AIA data into cubes with sizes of 9 and 5 TB, respectively. K. Dissauer et al. ([2023](#)) uses an extended HMI AR (HARP) cutout that has been identified in the photosphere and tracks the extended area of those ARs at a 12 minute cadence, allowing users to access EUV images of those ARs at the same cadence as SHARPs are calculated. On the other hand, R. Galvez et al. ([2019](#)) down-sample AIA images into a 512 × 512 image sampled at a 6 minute cadence. The first provides AIA patches constrained to the vicinity of the HARP boundary boxes instead of full-disk images, while the second is a condensed data set that down-samples images into one-eighth of their original size. Both provide a much-needed product that will make AIA's petabytes of data accessible to the scientific community. None exploits the full capabilities of the AIA.

## 3. Data and Motivation

Here, we describe how solar flares are defined and establish some limitations to current flare definitions, while also demonstrating the lack of spatial localization of canonical flares (Section [3.1](#)). We describe how differential analysis of the full-disk XRS signal is used as a rudimentary approach to identify C-class flares and above in Section [3.2](#). Finally, in Section [3.3](#), we establish and describe the sample of >1000 flares that the ALEXIS pipeline will locate in this proof-of-concept paper. For a discussion of the SQLite databases referenced in these sections, refer to Appendix [C](#).

### 3.1. Current Catalogs Do Not Know where More than 50% of Flares Occurred: The Flare Location Problem

Recall that the SWPC catalog is not perfect at identifying where on the solar surface the flare they reported occurred. The XRS instrument is basically a single-pixel detector that identifies all of the X-rays emitted from the Earth-facing side of the Sun at a given moment. Data compiled from the HEK show that the SWPC fails to locate solar flares for more than 50% of the events in their catalog (Figure [1](#)(A)).

It is also important to understand how solar flares are detected. The X-ray class or label of a solar flare is specific to the date and time when the flare peaked. Regardless of X-ray magnitude, the SWPC office, which is part of the NOAA, uses a universal metric to detect, catalog, and characterize solar flares. The algorithm used to define solar flares uses data points that represent the 1 minute average of the X-ray flux, as seen by the 1–8 Å or B wavelength of the XRS instrument. The technique that the SWPC group uses to identify the start, peak, and end of a flaring event as reported by the GOES website[4] is the following:

> "The event starts when 4 consecutive 1 minute X-Ray values have met all three of the following conditions: (a.) All 4 values are above the B1 threshold and (b.) All 4 values are strictly increasing and (c.) The last value is greater than 1.4 times the value which occurred 3 minutes earlier. The maximum is the time when the flux value reaches maximum. The maximum flux value (the event size) is the flux, as defined by the C-M-X scale, at the time of maximum. The event ends when the current flux reading returns to 1/2 the "peak" (peak is the sum of the flux at maximum plus the flux value at the start of the event)."

This way of defining a flare is problematic for the following reasons: (1) the GOES satellite has a 2 s cadence, and using the 1 minute average causes the removal of important data; (2) the X-ray background can easily be above the B-class flux for hours before the flare occurs, causing the algorithm to flag a flare when prolonged contamination occurs; (3) there is no physical evidence pointing toward the use of a flux greater than 1.4 times the value of three previous data points; and (4) the end of the flare is defined when the flux returns to half the maximum height, which assumes that the full-disk X-ray flux is the product of only one flaring region. It should be noted that the current operational pipelines that flag and filter possible eruptive events within the XRS data have been updated for the GOES16–18 iteration (see the GOES 16 read-me file).[5] The read-me file also describes a quad-diode

---

technique to use the new Solar Ultraviolet Imager (SUVI) data to localize flares. The SUVI is an EUV imager that replaced the soft X-ray images taken by SXI in the new generation of telescopes. The filtered images from SUVI closely follow the AIA band passes (i.e., 94, 131, 171, 195, 284, and 304 Å). There is room for improvement in both techniques, demonstrated by examples of multiple events, as shown in Figures 12, 13, 14, 15, 16, 17, 22, 23, and 25. Regardless, a data product with solar flare locations and a census independent of current techniques and constraints created with the XRS data from GOES13–15 was missing. This data product is important because the time span the GOES13–15 telescopes operated overlaps with the era of high variability and resolution studies solely enabled by the SDO. ALEXIS will provide such a data product. For example, the first technique we will discuss is used to flag possible flare peak times from the soft X-ray data and is dependent on the simple technique of differential analysis of the XRS signal.

### 3.2. Differential Analysis of the GOES XRS: Independent Verification of Reported Flare Peak Times

We have described above how the SWPC detects a flaring event using the single-pixel XRS data and its limitations. The purpose of this section is to describe how we use a differential analysis of the XRS signal to circumvent these limitations. The differential analysis of the XRS data can reproduce the date and time of canonical solar flare peak times. We describe how this technique works and the implications of its results.

ALEXIS's differential technique is composed of four parts (see Figure 1(B)). First, we query our XRS database for every hour of data from March 2010 to May 2020 and filter the queried result to only include XRS-B data (Appendix C). XRS data are returned at the minimum ∼2 s cadence. Second, we run a linear B-spline fit (P. Dierckx 1975) for the data from every instrument available in that hour; if there are two instruments available (i.e., GOES15 and GOES13), we will fit their 2 s time series individually for both their long and short bandpasses. Third, we resample the spline fit at a 3 minute cadence. Lastly, the derivative between these equally sampled points is taken, and the slope of the derivative is analyzed. We flag as a possible flaring event the moment in time where the derivative changes slope.[6] Specifically, when the derivative crosses from positive to negative, it typically corresponds to a local maximum in the signal. Thus, the zero-crossing point serves as a marker for potential local maxima, helping to identify peaks in the XRS signal that may correspond to flaring events.

We can compare these results from the differential technique with the peak times reported for events from the databases produced by the SWPC and SolarSoft and reported to the HEK. We employ an aggregation function similar in concept to a $k$-means approach. This approach places all peak times identified by each team on a list ordered by time. If multiple GOES telescopes are available and all report local maxima within 10 minutes of each other, we first aggregate the time stamps into a single local maxima at the mean of the GOES events. A pointer will parse through that list, finding all values within

5 minutes' worth of data, and encoding the number of teams within that time range.

ALEXIS's differential approach is not free from errors. For example, this simple technique will miss quantifying flares that are cotemporal but not related spatially. It is worth noting that this technique will also yield many false positives due to moments in time of rapid variability caused by noise in the XRS signal. Note that we are not claiming that the differential technique has found flares but rather using the local maxima as motivation to search for them with multispectral solar imagers. Nevertheless, the differential is sufficient to recover all the flares that the SWPC and the SolarSoft team identified as detected using the XRS sensor on board GOES13–15 (see Figure 1(C)).

### 3.3. Proof-of-concept Solar Flare Sample: 13 TB of Images

This proof-of-concept paper will concentrate on a sample consisting of 1000 time ranges centered at a local maxima that the SWPC, SolarSoft, and ALEXIS agree on (center of the Venn diagram in Figure 1(C)). Nevertheless, there might be other flares reported within the time ranges used for this analysis. As such, there will be more than 1000 flares that the ALEXIS pipeline needs to recover (see Figure 2).

We are interested in understanding the high resolution and cadence in the EUV and X-ray range for a large sample of flares. Downloading the data locally is initiated with a datetime object. For each local maxima ($Q$), ALEXIS queries all available data for $Q \pm 20$ minutes from the AIA availability database and $Q \pm 40$ minutes from the SXI availability database. The time ranges for each instrument were chosen such that we can save as much disk space as possible while having sufficient data points for analysis. In general, ∼1500 files will map to each $Q$ (i.e., $\frac{l_{image}}{12_s} \times 40$ minutes ≈ 200 images per AIA wavelength (94, 131, 171, 193, 211, 304, 335 Å), and for the SXI there are a maximum of $\frac{l_{image}}{5_{minutes}} \times 80$ minutes ≈ 16 for each band pass (TM, PTHN, PTHK)).

In order to ensure that images are not downloaded twice, the URL path specific for each file is passed into the cryptographic hash function called Secure Hash Algorithm 256-bit (SHA-256). This guarantees that each input message (URL) will map to a unique deterministic 256-bit hexadecimal string. Only unique hashed URLs are downloaded using wget,[7] and the output log is saved in a different directory. The output logs are parsed using regular expressions to cache the speed and size of the downloaded instance. If the download speed and size cannot be found, the files are deleted and deemed either corrupt or rate limited by JSOC or NOAA. Once the download pipeline finishes, if you were to run it again, it would run only files that are not present on the target file system. This allows corrupt or rate-limited files an extra chance to be downloaded correctly; every time the download script is run, ALEXIS will try to download and parse only the corrupt or missing files once again. All files are stored and organized in a directory named after the first two letters of the SHA-256 keyword. Each file is queried as needed during the main ALEXIS pipeline.

One detail that makes the results of this paper possible is that the ALEXIS pipeline requires the highest resolution and cadence of multipixel images available for heliophysics; none of the downloaded files are downsampled in time or space during the analysis. In total, 1,505,151 unique images that occupy ∼13 TB of disk space were analyzed for this proof-of-concept paper. To the authors' knowledge, this proof-of-concept run represents the largest use of AIA data in the literature. In a second paper, ∼14 million files that require upward of 100 TB of data will report on verifying more than 8000 solar flares from 2010 March to 2020 May (J. R. Padial-Doble 2025, in preparation). The analysis for the second paper has been completed and is being validated.

---

[6] It is critical to have equally spaced data points in order to do a differential analysis. For a signal $S$, the differential $\frac{dS}{dt}$ will only be valid if the signal is taken at equal intervals of time ($\triangle t$). Assuming that the raw XRS data are taken at equal intervals and taking the derivative at the Nyquist sampling frequency is a poor simplification of the problem.

[7] https://www.gnu.org/software/wget/

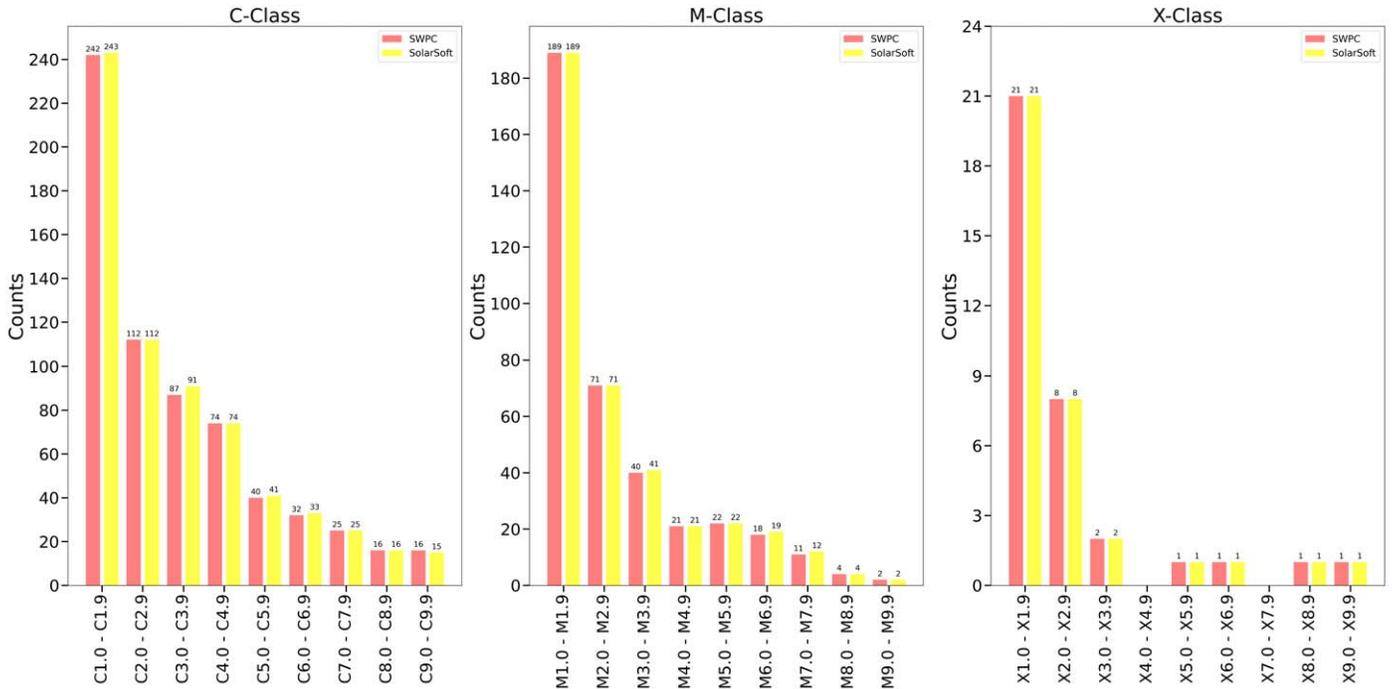

**Figure 2.** Distribution of flares by GOES class that ALEXIS is tasked with finding; 1000 random time ranges are chosen. The SWPC catalog returns 648 C-class flares, 378 M-class flares, and 35 X-class flares (1061 flares in total). SolarSoft returns a total of 653 C-class flares, 381 M-class flares, and 35 X-class flares (1069 flares in total). Note that all reported X-class flares between 2010 May and 2020 May are included in this iteration of the ALEXIS pipeline. In general, we assume SolarSoft uses the flares produced by the original SWPC catalog. Upon visual inspection, SolarSoft reports 11 false positives and the SWPC reports two false positives.

However, not all data are of science quality or usable (see Appendices A and B). Once the quality of each image is assessed, only wavelengths with a minimum number of files across a given time range and continuous in time are passed to the pipeline. For example, let us assume there are three wavelengths that have data from a given time range centered on a local maxima time stamp: AIA-193, AIA-131, and SXI-GOES15-TM. Assume good data quality for 183/200 images from AIA-193, 199/200 images from AIA-131, and 10/16 images from SXI-GOES15-TM. Assume also that the 17 images from AIA-193 that are not science quality were the last 17 images of the time range.

Note that the AIA-131 data have more than 90% of the possible images that are science quality and there are no large chunks of missing data (i.e., only missing 12 s from 40 minutes of data). AIA-193 has more than 90% of the possible images that are science quality, but we are missing 3 minutes and 24 s of continuous data (i.e., 17 chronological images missed). Only 63% of science-quality images are reported for SXI-GOES15-TM. This result is only that the AIA-131 wavelength will be passed onto the pipeline. For data to be passed onto the pipeline, there must not be a chronological set of time stamps where the data are missing. AIA-193 and SXI-GOES15-TM will not go through to the pipeline.

## 4. Methods

After introducing the ALEXIS data products (Section 4.1), this section describes how the pipeline works. The technique employed by ALEXIS to identify and localize flares is described in Section 4.2. The pipeline then resamples and organizes the data into a set of grid-searched Cartesian product-training matrices (Section 4.3). Then, a convex optimizer is tasked with minimizing the error between a weighted linear combination of candidate regions and the XRS signal (Section 4.4) where several evaluation metrics are calculated to choose the best regularizing value (Section 4.5). Finally, ALEXIS assigns an X-ray magnitude while spatiotemporally attributing a solar flare to a HARP (Section 4.6). For a more complete discussion of the data-cleaning techniques used, please refer to Appendices A and B. For a discussion on the SQLite databases referenced in these sections, refer to Appendix C.

### 4.1. Overview of ALEXIS Data Products

The ALEXIS pipeline provides two data products and one software package. The first data product is an event list, complete with flare metadata such as peak time, GOES X-ray class, NOAA AR number, Helioprojective Cartesian (HPC) $x$-coordinate, HPC $y$-coordinate, HPC boundary polygon, ID team, and search instrument, as well as the HARP number for the flares from all teams. The second data product is a light curve for each available wavelength created by the optimized fit between the X-ray and all the EUV/soft X-ray images. This will allow us to study the time differences between the peak emission from multiple EUV/soft X-ray and the XRS data, which could provide information about cooling rates of plasmas to these imaging temperatures. In addition to the bulk analysis of a solar data set, users can isolate specific 1 hr time segments to facilitate external validation by providing a specific date-time and asking the ALEXIS pipeline to query, download, parse, optimize, and return the fitted light curves. Figure 3 depicts the end products of the pipeline; all data to generate these figures and the infrastructure to manipulate the data are provided on ALEXIS GitHub.[8]

Appendix D shows a variety of examples of the ALEXIS pipeline, highlighting its capabilities. We include examples of flare locations reported by canonical databases that are incorrect but recovered by ALEXIS (Figures 12, 14, 16, 17, 20, 22, and 24), examples where the flare duration reported by canonical databases are incomplete (Figures 12, 13, 14, 15, 16, 17, and 25), examples where the X-ray magnitude for a flare is insufficient to describe the overall strength of the flare emission (Figures 14, 16, 17, 19, 22, 23, and 24), examples where ALEXIS detects the background emission before or after a flare (Figures 18 and 19), examples where there are two flares happening within the same AR (Figures 20 and 21), examples when multiple ARs are flaring at the same moment (Figures 25, 22, 23, and 24), as well as segments when ALEXIS fails (Figures 14, 15, and 17).

### 4.2. Identifying Candidate Regions from AIA and SXI Images

The pipeline is initiated by giving ALEXIS a single date-time or multiple date-times corresponding to a flare peak (see Section 3.2). ALEXIS will then query and download all images within ±20 minutes for the AIA data and ±40 minutes of the SXI data (see Section 3.3). Within these images, a list of potential flare candidate coordinates is identified using two main techniques. Broadly, ALEXIS identifies interesting regions by first finding the brightest pixels in every AIA and SXI image. Then, ALEXIS clusters these bright regions in space, in time, and hyperspectrally into groups that we call candidate clusters or candidate flaring regions. One or multiple candidate flaring regions are returned for each time range under analysis. Figure 4 shows a cartoon that depicts the whole process of identifying possible flaring regions.

In more detail, ALEXIS first uses *peak_local_max* from Scikit-Image (S. van der Walt et al. 2014) to find all pixels that are greater than 90% of the local maximum (see the first row of Figure 4). This threshold is conservative enough to catch C-class flares. The peak finding is performed for every image that passes the cleaning procedure described in Appendices A and B. Each image will always return at least one pixel that satisfies this threshold. After all the bright pixels are found, we employ a filtering process to choose which groups of pixels will be evaluated in the rest of the pipeline.

ALEXIS's filtering process is composed of three density-based spatial clustering of applications with noise (DBSCAN) methods from Scikit-Learn (M. Ester et al. 1996; F. Pedregosa et al. 2011). The benefit of DBSCAN is that the total number of clusters is not predetermined. However, there are two main hyperparameters that must be specified: the $\epsilon$ neighborhood of a point and the minimum number of members within $\epsilon$, $N_{\mathrm{member}}$. The $\epsilon$ neighborhood (or "eps") is the maximum distance needed to distinguish core points from outlier points, while the integer $N_{\mathrm{member}}$ establishes the minimum number of pixels needed to define core points and therefore constrain the pixels that belong to an individual cluster. We adopt an $\epsilon = 50''$ for each step in the algorithm. The other main hyperparameter, $N_{\mathrm{member}}$, will depend on the goal each implementation of DBSCAN aims to achieve.

The goal of the first iteration of DBSCAN is to parse through each individual image and join multiple pixels that are greater than 90% into one or more clusters (see the second row of Figure 4). This spatial clustering is achieved by setting $N_{\mathrm{member}} = 1$ and maintaining $\epsilon = 50''$. Note that, in the first image tiles for the wavelength 94 Å, many peaks are identified, some comprised of several members, and others are singular. The mean coordinates of all members of a cluster define the center coordinates of a spatial cluster. At the end of this spatial-clustering step, each image will have at least one independent coordinate pair.

However, since flares are not instantaneous, we require that these spatial clusters persist in time. This is the goal of the second DBSCAN. This temporal clustering requires that spatially defined clusters be bright for at least 6 minutes in the AIA or 15 minutes in the SXI (see the third row of Figure 4). Note that the flare magnitude relates to various physical measurements of solar flares such as the temperature and emission measure, but J. W. Reep & K. J. Knizhnik (2019) concluded that the flare duration and flare magnitude are independent of each other. Our temporal threshold was chosen by sampling all flares reported by the HEK and plotting a histogram of the flare duration as a function of flare magnitude; 6 minutes seemed to be a good first-order estimate. Centered at the differential peak time, the AIA's 12 s cadence over ±20 minutes results in 200 images per wavelength and the 5 minute cadence across ±40 minutes for the SXI results in 24 images. The minimum number of spatial core points within an $\epsilon$ of $50''$ must hold for 15% of the available images for any given wavelength. Given this information, the hyperparameter $N_{\mathrm{member}}$ will be chosen depending on which imaging instrument is undergoing analysis. For example, when ALEXIS uses AIA data, a minimum of 30 images of the same wavelength must have ≈6 minutes of spatial cluster within $\epsilon$. On the other hand, temporal clustering of SXI images must have a minimum of $N_{\mathrm{member}} = 24_{\mathrm{images}} \times .15 \approx 3$; the three images from the SXI track 15 minutes of spatial cluster within an $\epsilon$ of $50''$. Regardless of the imaging instrument, the centroid coordinates for each temporal cluster will be defined by the mean HPC coordinates of all the members that represent the temporal clusters for each wavelength.

The last implementation of DBSCAN requires that a flare occurs across multiple wavelengths (see the last row of Figure 4). The goal of this hyperspectral clustering is to filter and discard temporal clusters that occur only in a small number of wavelengths. Hyperspectral clustering is important because different physical phenomena occur from the photosphere all the way up to the corona. For example, the EUV and X-ray bands encode information from reconnection outflows, current sheets, fast shocks, magnetic footprints, evaporating plasma, and postreconnection hot loops (K. Shibata et al. 1995). Hyperspectral clustering joins the temporal clusters of every wavelength/instrument and returns the final candidate clusters. Hyperparameters are chosen such that at least three-quarters of the analyzed wavelengths have temporal coordinates within ALEXIS's standard DBSCAN hyperparameter $\epsilon$. The pixel coordinates where a flare might have occurred are the mean HPC coordinates from all members in the hyperspectral cluster.

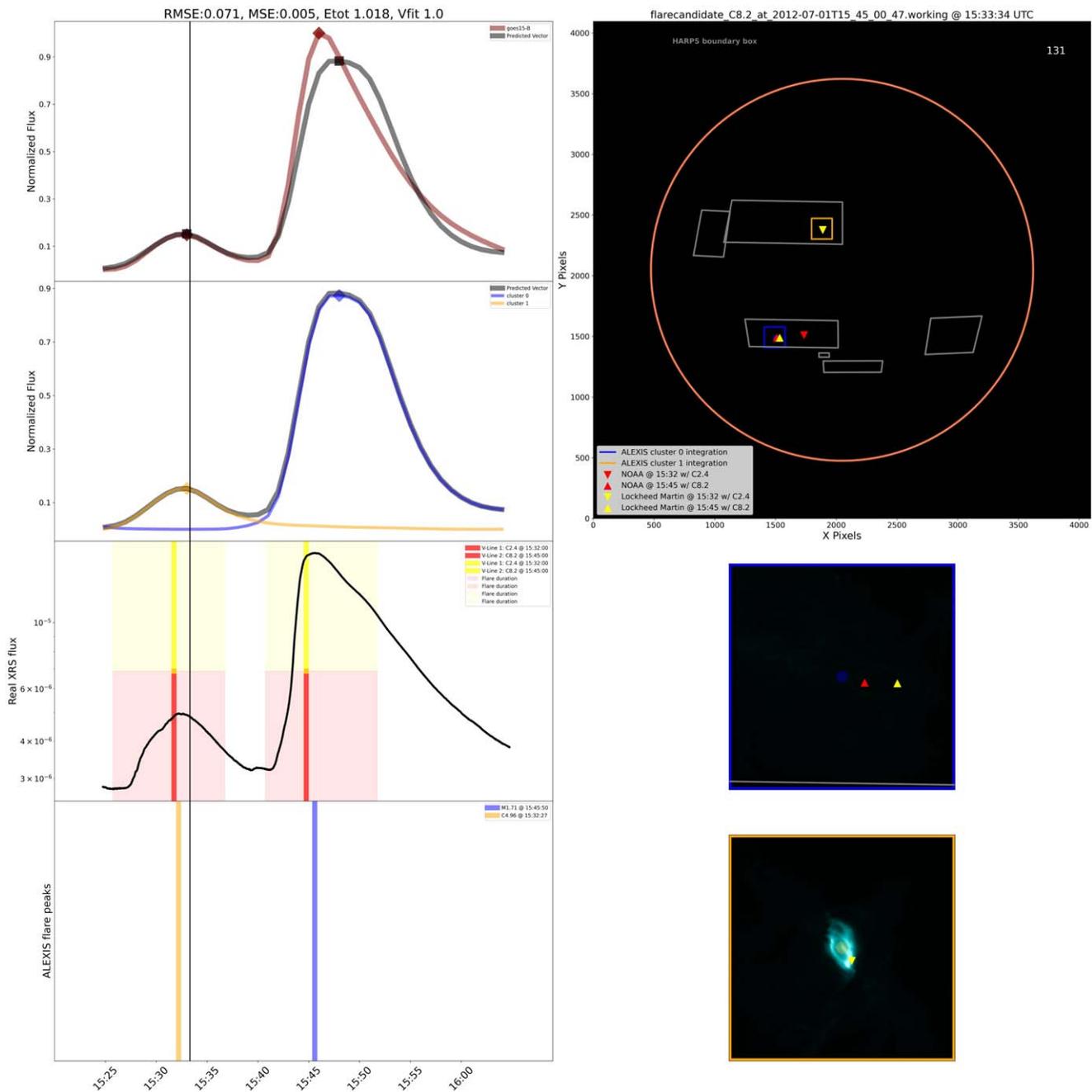

**Figure 3.** Visualization of convex optimization. Left: title will include the evaluation metrics for the subset. The first row shows normalized flux vs. time, where light red represents the X-ray flux integrated across the whole solar surface while the vector created by the linear sum of discrete regions is shown in gray. Note the black squares and red diamonds that mark the peaks of the signal, which are necessary to define peaks, as described in Section 4.5. The second row shows normalized flux vs. time where the time series that compose the linear combination vector are traced in colors that follow the integration regions. Blue is reserved for cluster 0, orange for cluster 1, green for cluster 2, and purple for cluster 3. Note the diamonds color coded to the cluster regions and indicate the peaks in the time series used to define the peak time of a flare. The third row shows the 2 s XRS-B data (black solid line) vs. time where the flare metadata for SolarSoft (yellow) and the SWPC (red) are overlaid on top. Each team reports a flare peak time (solid vertical lines) and the duration of the flare (lighter-colored vertical span). Note the label on the third column shows the GOES flare class and peak time for each respective team. The fourth row shows the peak times defined by ALEXIS color coded by the cluster region that produced them, and the legend provides ALEXIS's GOES X-ray class and peak time. The first row shows an image of the Sun at the wavelength that returned the best convex fit. The salmon-colored circle delineates the solar border, gray boxes show the HARP ARs available, and the colored boxes show the integration region of each cluster. Note the yellow and red triangles that show the location metadata as compiled by SolarSoft and the SWPC. If more than one flare is present in the canonical metadata for this time range, the triangle apex will be organized as follows: (upward) for the first flare in time, (downward) for the second flare in time, (left) third flare in time, (right) fourth flare in time. Below the main solar image, a zoomed-in region for each integration box is shown. The cluster the zoomed-in region belongs to is identified by the color of the border of the zoomed-in cutout. A colored dot representing the pixel returned by ALEXIS where the flare occurred is centered within the integration box. Specificity and accuracy: SolarSoft is accurate and specific for both entries. The SWPC is neither accurate nor specific in the first flare (orange) but is accurate and specific in the second flare (blue). See Section 5.2 for a discussion on accuracy and specificity.

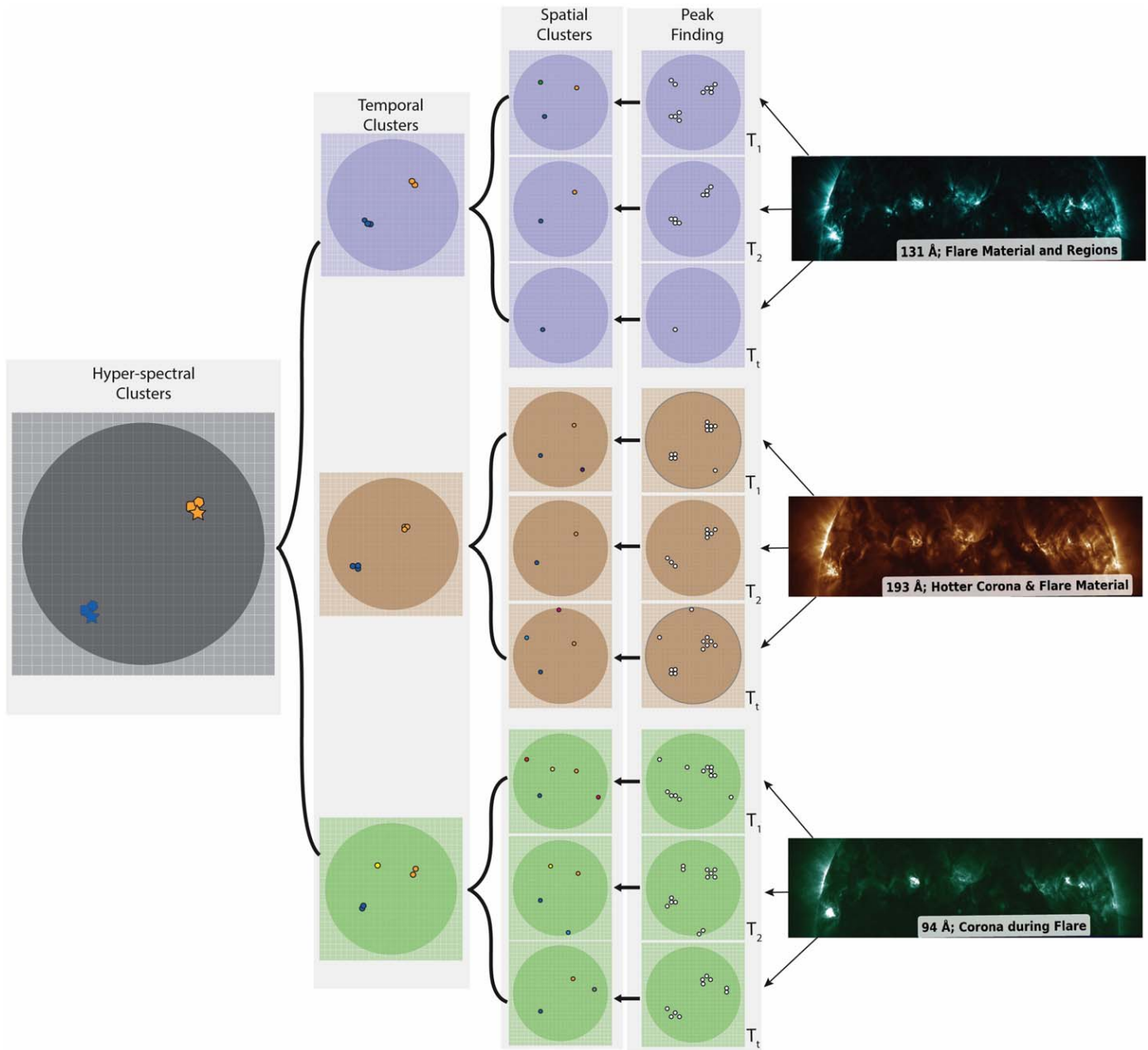

**Figure 4.** Identifying candidate flaring regions: for every science-quality image, ALEXIS finds all pixels that are 90% of the maximum of that image. The first density-based clustering is applied to each image. Spatial clusters are designed to return the mean coordinates of a cluster with a minimum of one member located within the radius of 50″. The second clustering collapses the results of spatial clusters in the time dimension. Temporal clusters are designed to find regions that are bright for at least 15% of the total amount of images available within a radius of 50″. Finally, the wavelength dimensions are collapsed and we begin hyperspectral clustering. This will return the coordinates of long-lasting bright regions that occur in at least three-quarters of the available wavelengths. This process may return one or more candidate flare coordinates.

From the list of candidate regions that have survived spatial, temporal, and spectral clustering, ALEXIS calculates the integrated flux over a bounded region centered at the candidate coordinates. The main goal of generating these integrated flux time series is to find the best combination of candidate regions that can recreate the observed X-ray flux from the full solar disk. To do so, we create matrices that represent all available combinations of regional flux by employing a process called grid-searched Cartesian product, described in the next subsection.

### 4.3. Grid-searched Cartesian Product: Creating Training Matrices

We are interested in tracking the emission around the possible flaring locations uncovered by spatial, temporal, and hyperspectral clustering, so we generate time series of the flux from each location. ALEXIS successively "zooms in" on each of the candidate clusters to identify the smallest possible boundary box, which excludes emission that is unrelated to the flare. Integration boundary boxes are of areas no greater than 150, 100, 80, and 50 arcsec$^2$, generating four separate time series for each candidate flaring region. These time series are then fit with a linear B-spline interpolator from the Scipy package (P. Dierckx 1975; P. Virtanen et al. 2020) and resampled at a minute cadence to produce a coincident data stream. The minute cadence chosen here is conservative. We could, if we wanted, sample at a Nyquist frequency of 24 s.

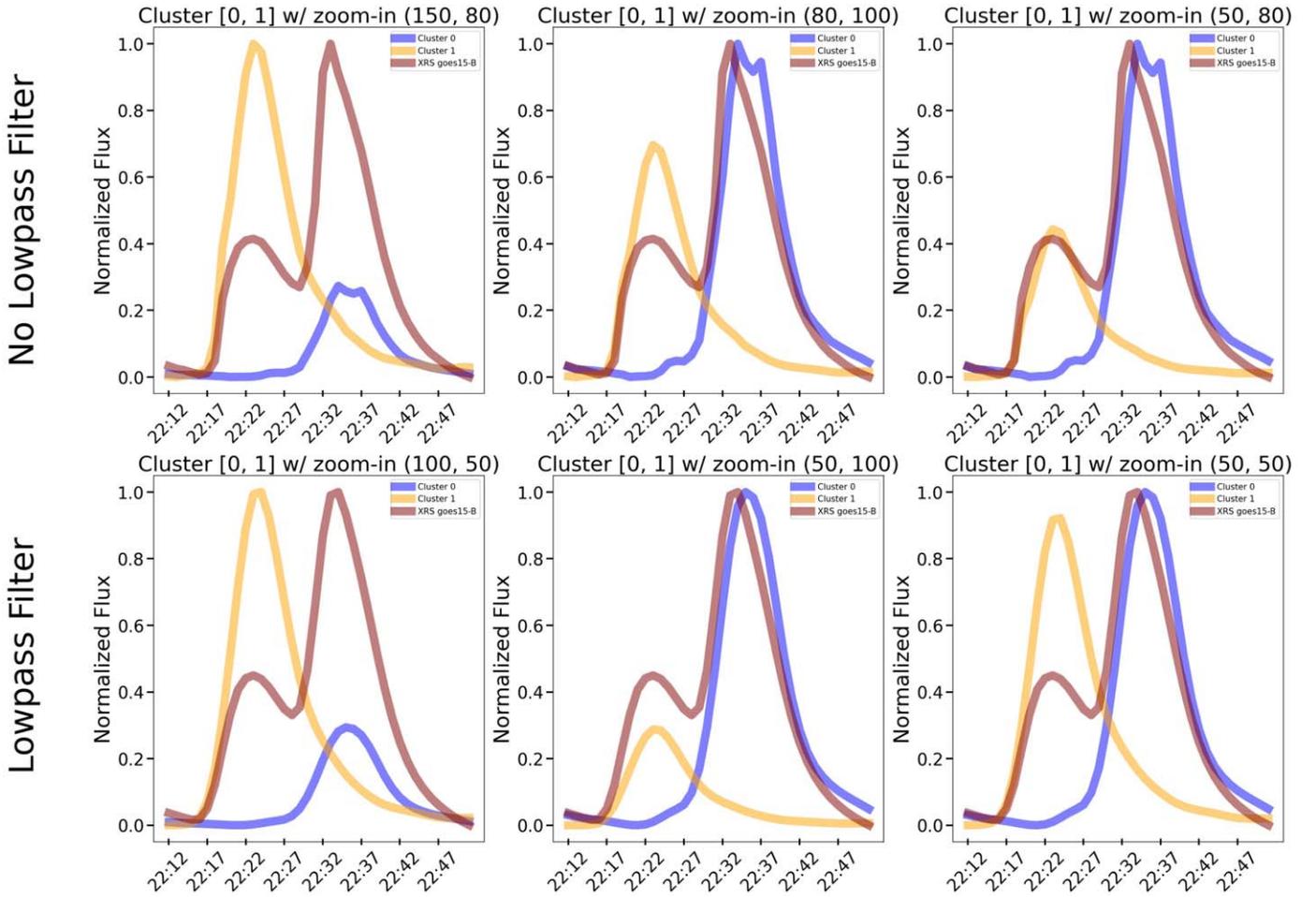

**Figure 5.** Data representation of a grid-searched Cartesian product: sample of training matrices for six zoomed-in Cartesian products and the grid-search tuple of two regions. ALEXIS searches for a C7.4-class flare that occurred at 22:31 UTC on 2014 January 7. Training matrices as seen by AIA-131 Å (L) paired to the GOES15-XRS-B target signal (Z). Each plot shows the normalized flux as a function of time. The title of each plot shows the cluster tuple with the specific zoom-in for each cluster. Maroon (XRS data), blue (integrated flux for region 1), orange (integrated flux for region 2). Upper row: no low-pass filter applied. Changes in exposure time predetermined by the AIA cause the signal to change value quickly. The signal is not smooth. Lower row: low-pass filter applied. The effects in the integrated flux from saturated pixels and nonsaturated pixels affected by the exposure time changes are filtered out; the signal is smoothed but the overall timing of peaks and valleys are retained. ALEXIS identifies two flares: a C8.4 flare happens at the orange candidate region and an M1.5 flare occurs at the blue candidate region.

ALEXIS now has multiple time-series vectors: a vector representing each zoomed-in integrated flux for every candidate region at every imaging wavelength, and a vector representing each available XRS wavelength for every available GOES instrument. For uniformity, the minimum of the signal is subtracted from each time-series vector. This process is called offset or direct current correction. Subtracting the minimum value recenters the data to a common baseline for normalization, removes bias or noise related to the observation instrument, and focuses our analysis on the general changes in trend of the time series instead of the absolute magnitudes. Potential pitfalls from using this technique include the attenuation of smaller flares in the presence of a significantly larger one.

The pipeline then combines the flux of separate candidate regions in two steps to create training matrices for the linear convex optimizer. First, a grid search over the candidate regions is created. Consider the example in which ALEXIS found three distinct coordinate pairs labeled region 1, region 2, and region 3. A grid search will produce a list of the following combinations: [(region 1), (region 2), (region 3), (region 1, region 2), (region 1, region 3), (region 2, region 3), (region 1, region 2, region 3)]. In total, a grid search on $N$ possible candidate locations will produce $2^N - 1$ tuples. Next, ALEXIS creates matrices that represent all possible combinations between the grid-searched candidate regions and each zoomed-in image.

To outline this in greater detail, assume that two possible flaring locations were found, and we focus specifically on the grid-search tuple that corresponds to both regions being used (i.e., $N = 2$; (region 1, region 2)). ALEXIS will take this tuple and create a $4^N$ training matrix combining all possible zoomed-in fluxes for each region. For example, the components for one of the matrices created will represent the first region with an integrated boxed area of 50 arcsec$^2$ combined with the signal created by the second region integrated over a 150 arcsec$^2$ box. Another matrix contains the first region integrated over the 150 arcsec$^2$ box, the second region with the 80 arcsec$^2$ box, and so on. This process continues until each possible combination of zoomed-in integration areas for every candidate region within that specific grid-searched tuple is satisfied. The process of finding all possible combinations of zoom-ins while maintaining the order of the candidate regions is called a Cartesian product.

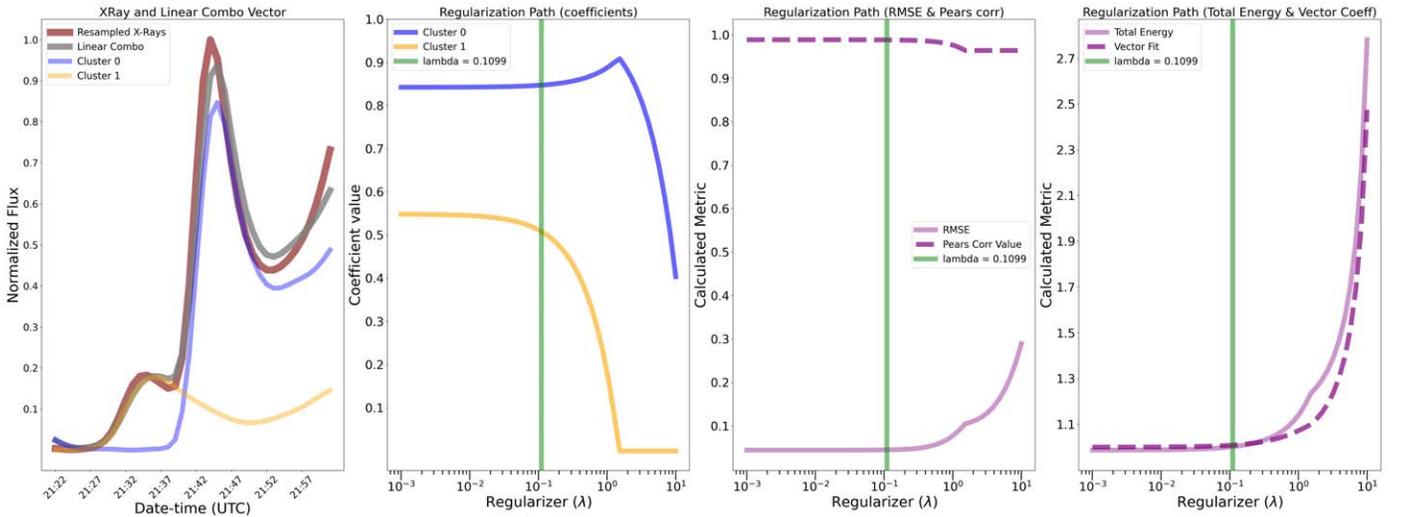

**Figure 6.** Regularization path: we are interested in LASSO regression to eliminate candidate regions that are unnecessary. Column (1): X-ray and linear combination vector. Light red follows the target XRS signal with light orange and light blue, showing the individual weighted flux from the candidate regions. In gray, we show the predicted vector. Column (2): regularization path. Light orange and light blue show the value of each respective coefficient across the regularization path. The green vertical line indicates the $\lambda$ value that produces the coefficients that created the predicted vector (gray); the green vertical line will progress through the values of $\lambda$ from $10^{-3}$ to 10. Columns (3)–(4): the value of different evaluation metrics at each point in the regularization path. For each value of $\lambda$, we compute differences between the predicted vector ($\hat{Y}'$) and the XRS target signal ($\hat{Y}$). Here, we show how the Pearson correlation coefficient, rms error, mean square error, a least-squares model over the predicted signal, and its total energy differs as $\lambda$ increases.

The grid-searched Cartesian product described above is performed for every wavelength and telescope pair of imaging data available. ALEXIS has created a library of ($N_{ijk} \times T$) matrices where $T$ is the number of resampled time steps and $N_{ijk}$ is the row representing the $i$th possible candidate clusters integrated using the $j$th squared arcsecond boundary box for the $k$th available wavelength-telescope pair. In total, $L$ number of training matrices are produced: $L = (5^N - 1) \times W$ where $W$ is the number of unique wavelength-telescope pairs available for analysis. In the two-region example above for 12 wavelengths, $L$ is 372.

It is important to note that there is noise in each of the $L$ matrices described above. This noise can be a product of the rapid changes in the integrated values from one time step to the other or because the numerical limits of each detector have been reached. A zero-phase low-pass filter (F. Gustafsson [1996]) is applied to deal with rapid changes in the time series, ensuring that the general trends in the signal are preserved. This technique allows the time series to not be shifted in time, so that the true timing of peaks in the signal is conserved. In general, zero-phase low-pass filtering will smooth out the high-frequency changes in the signal. The values within each matrix are then normalized to unity dividing by the largest value found in the matrix.

Each matrix in $L$ is then paired to a target signal. The target signal is the resampled XRS data stream that has also been offset corrected, low-pass filtered, and normalized to unity by dividing with the maximum emission during that time range. In other words, the (1) resampled and normalized time series for the long and short band-pass XRS signal will be paired to (2) every training matrix created by the grid-searched Cartesian product-integrated flux. This combination creates $L \times Z$ subsamples, where $Z$ is defined as the combination of primary or secondary GOES telescopes in operation at the moment and their respective long and short bandpasses. Subsamples are defined as the specific combination of data composed of XRS wavelength (A or B), XRS telescope (GOES13, GOES14, GOES15), image telescope (SXI or AIA), image wavelength (94, 131, 171, 193, 211, 304, and 335 Å, TM, PTHN, and PTHK, grid-searched tuple, and Cartesian product zoomed-in type. Figure [5] shows the effects of low-pass filtering on six subsamples.

The resampled, normalized, offset corrected, and zero-phase low-pass filtered time series of both the candidate flare matrix and the XRS vector that define each subsample are passed onto a convex optimizer. The convex optimizer is tasked with minimizing the difference between an XRS vector and a weighted training matrix. The dot product will produce a single vector that represents the weighted linear combination of each column in the candidate region matrix. In the next section, we will describe the least absolute shrinkage and selection operator (LASSO) minimization algorithm used by the convex optimizer.

### 4.4. Filtering Candidate Flaring Regions: LASSO Regularization

To what degree can the linear combination from the flux of each individual region resampled from the EUV or soft X-ray's images recreate the resampled single-pixel full-disk X-ray flux? We approach this question using a LASSO minimization algorithm (see R. Tibshirani [2011], and references therein). The goal is to find the combination of coefficients for each column of a training matrix such that the linear combination of the integrated flux most closely resembles the flux from the XRS. The evaluation of the goal requires minimizing the square difference between the fitted signals in the presence of a penalty term. The degree in which the penalty is applied will vary according to the square root of the magnitude of the coefficient vector returned by the convex optimizer. LASSO is chosen because of its ability to promote sparsity in the coefficients; LASSO allows the weighted coefficients to be zero in contrast to a ridge regression algorithm. The optimized set of coefficients are constrained between [0, 1] and interpreted as the relative importance of the flux for that region. In other words, a value of 0 for the coefficient of one region requires no signal from that cluster, a value of 0.10 for a coefficient will require 10% of that signal, and a value of 1.0 implies that the full normalized signal from that cluster is vital.

LASSO can be described as the minimization of

$$\arg\min_{\beta}(||\beta\hat{X}^T - \hat{Y}||_2^2 + \lambda||\beta||_1), \tag{1}$$

where $\hat{X}$ is a $(T \times N_{ijk})$ matrix created by the grid-searched Cartesian product that has been zero-phase filtered. The single-pixel X-ray vector $(T \times 1)$ is $\hat{Y}$ that has also been offset corrected, zero-phase filtered, and normalized to unity to the maximum emission for that time range, and $\beta$ is a $(1 \times N)$ vector whose members are the rational coefficients constrained between [0, 1]. The dot product between $\beta$ and $\hat{X}^T$ will create a $1 \times T$ vector that we will call the predicted vector $\hat{Y}'$. The task optimizes the set of weights in $\beta$ that minimizes the error between a predicted vector $(\hat{Y}')$ and a known target vector $(\hat{Y})$ while incurring some degree of penalty, called the $l_1$ penalty.

The degree to which we penalize LASSO depends on the range of values that the fit allows the penalty $(\lambda)$ to take and the square root of the magnitude of the $\beta$ vector. The higher the value of $\lambda$, the more the linear fit is penalized. We allow $\lambda$ to take on a wide range of 50 logarithmically sampled values between $10^{-3}$ and $10^1$. For every $\lambda$, a new set of coefficients is fit, which gives rise to a regularization path that depends on the degree of penalty $(\lambda)$ applied. This is shown in the first two columns of Figure 6. LASSO minimization was written using the convex optimization library CVXPY (S. Diamond & S. Boyd 2016) and constrained to optimize individual elements of the $\beta$ vector between the values of [0, 1]. Evaluation metrics between the predicted vector $(\hat{Y}')$ created at each step in the regularization path and the XRS vector $(\hat{Y})$ are calculated.

### 4.5. Evaluation Metrics: Choosing the Optimal Regularizer (λ)

Regularization ensures that our solution is simple and generalizable. There are three ways in which we can choose each vector optimized in the regularization path. There are five calculated metrics that compare the optimized predicted vector $(\hat{Y}')$ with the X-ray ground-truth vector $(\hat{Y})$. The metrics included are Pearson's correlation factor (Equation (2)), rms error (RMSE; Equation (3)), mean square error (MSE; Equation (4)), total energy (Equation (5)), and vector similarity (Equation (6)). When comparing two time-series models with similar complexity, the standard RMSE and MSE are often preferred because they use the total number of observations $T$ in the denominator, providing a direct measure of prediction error without any adjustment for model complexity. These are simple methods to understand the raw deviation from the general normalized error.

$$r = \frac{\sum_{i=1}^{T}(\hat{Y}'_i - \bar{\hat{Y}'})(\hat{Y}_i - \bar{\hat{Y}})}{\sqrt{\sum_{i=1}^{T}(\hat{Y}'_i - \bar{\hat{Y}'})^2}\sqrt{\sum_{i=1}^{T}(\hat{Y}_i - \bar{\hat{Y}})^2}}, \tag{2}$$

$$RMSE = \sqrt{\frac{\sum_{t=1}^{T}(\hat{Y}'_t - \hat{Y}_t)^2}{T}} \tag{3}$$

$$MSE = \frac{\sum_{t=1}^{T}(\hat{Y}'_t - \hat{Y}_t)^2}{T} \tag{4}$$

$$E_{total} = \arg\min_{\alpha_{Etot}}\left(||\alpha_{Etot}\sum_{t=1}^{T}\hat{Y}_t'^2 - \sum_{t=1}^{T}\hat{Y}_t^2||_2^2\right) \tag{5}$$

$$V_{full} = \arg\min_{\alpha_{full}}(||\alpha_{full}\hat{Y}' - \hat{Y}||_2^2). \tag{6}$$

Once every subsample is optimized, we choose the $\lambda$ within the regularization path that produces the metrics (RMSE, MSE, Pearson's, $E_{total}$, $V_{full}$) closest to the target value of (0, 0, 1, 1, 1). In other words, if ALEXIS is able to recreate the XRS vector perfectly, the evaluation metrics that compare the predicted vector $(\hat{Y}')$ and the XRS vector $(\hat{Y})$ would lie within the five-dimensional plane delimited by (0, 0, 1, 1, 1). At each point in the regularization path, for every subsample, we calculate the Euclidean distance of the predicted vector from the optimal evaluation plane. The linear combination vector $(\hat{Y}')$ whose Euclidean distance (Equation (7)) to the perfect fit is closest to 0 in this evaluation space is chosen as the best multipixel representation of the single-pixel truth.

$$D = \sqrt{w_{RMSE} \cdot (RMSE - 0)^2 + w_{MSE} \cdot (MSE - 0)^2 + w_r \cdot (r - 1)^2 + w_{E_{total}} \cdot (E_{total} - 1)^2 + w_{V_{full}} \cdot (V_{full} - 1)^2}. \tag{7}$$

the optimal regularizer. The first case chooses the lowest regularizing value that describes a simple least-squares model without penalty. The second case would be to choose a constant regularization applied to each individual subsample. The third approach would be to choose the best-fit penalty by evaluating the regularization path; for every penalty, we can calculate some set of metrics and choose the penalty that best satisfies the intended question. We follow the third approach as described below.

In total, 50 attempts are made to recreate a subsample for every penalty value of $\lambda$, and metrics will be calculated for

For this first iteration of the pipeline, we chose equally weighted distances for all metrics except $w_r = \frac{1}{4}$. This weight for the Pearson correlation factor will always map between [0, 1]. We justify our decisions as follows: an RMSE = 1 implies that the two signals are completely opposite to each other, and the same is true about MSE. This implies that the Euclidean distances for RMSE and MSE values must be less than 1 and, thus, constrained. Now the optimal total energy and the vector fit linger around 1, but generate larger numbers when the fit is really bad. Taken together, this means that each term inside the distance calculation, in the context of a "good fit," is constrained to 1. The authors still tried several weighting methods and noted that changing the weight of each metric did not substantially change our results. We understand that there must be an optimal weight for each of the metrics, which will be evaluated in future work.

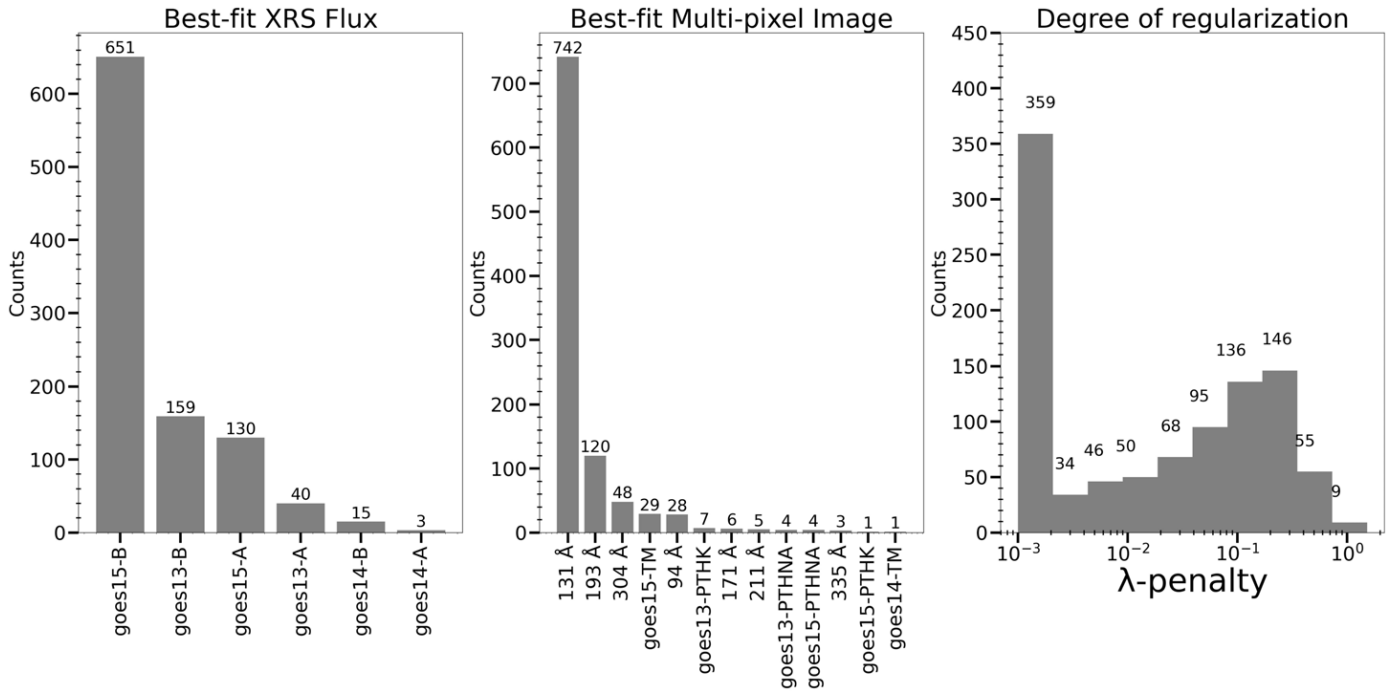

**Figure 7.** Which multipixel wavelengths and XRS telescopes return the best optimized fit? The single-pixel GOES15-B combined with the multipixel SDO 131 angstrom multipixel image returns the best fit. What degree of regularization is needed for this sample? Possible regularization values span 50 rational numbers sampled from a logarithmic space from $10^{-3}$ to $10^1$. Within the bin for $\lambda < 2 \times 10^{-3}$ lies 36% of the sample, indicating a small degree of regularization needed for one-third of the sample. The average $\lambda$ is $0.09 \pm 0.16$, suggesting that there is a large spread of regularization values. Regularizing our fit is a technique that produces more general models.

If the subsample closest to the origin has any coefficient <0.1, that particular candidate region $N_{ijk}$ is deemed unnecessary, and the optimized subsample closest to the origin without those candidate regions is chosen. This cutoff requires ALEXIS to choose only candidate regions where more than 10% of their signal is needed.

### 4.6. HARP Association, Flare Detection, and X-Ray Class

Now that ALEXIS has found discrete regions that contribute to the XRS signal, it associates and assigns the surviving candidate flaring regions with the relevant metadata. This metadata includes associating a HARP, determining which candidate cluster flared, assigning an X-ray label to the flaring candidates, and aggregating in space and time ALEXIS's candidate clusters to known flare information from canonical databases. This known metadata includes information from the NOAA team (the SWPC) and the Lockheed Martin team (SolarSoft) as reported to the HEK.

First, ALEXIS queries the HARP SQLite database and cycles through predefined time intervals until the HARP entries are found. These intervals span time ranges of 1, 2, 4, 10, 16, 33, and 46 hr. HARP boundary boxes are taken in the Heliographic Stonyhurst reported by the *hmi.sharp_720s* data series. All candidate flare regions from the optimized matrix are associated spatiotemporally with a HARP. Relevant metadata such as the HARP number, the coordinates of the spanned boundary box, and associated NOAA AR numbers are cached. Note that the *hmi.*

*sharp_720s* data series has been postprocessed to include the total boundary box in projected coordinates as seen once the HARP has progressed its transit across the solar surface ($\approx$14 days) but not in detector coordinates. In order to create the HMI SHARP boundary boxes in detector coordinates and projected onto the AIA image, we must know the HMI–AIA plate–scale ratios and the HARP CRPIX, IMCRPIX, XDIM_CCD, and YDIM_CCD keywords, which is beyond the scope of this work. Associating the flare to a HARP in projected coordinates is a more difficult problem to solve since, in the detector coordinates, the HARP areas are extensively larger than in projected coordinates.

Next, ALEXIS begins to define a flare by running a peak finder over several time series and aggregating peaks that are close in time. The XRS vector, the predicted vector, and the individual candidate region's vector must have peaks in their signals that coincide within 2 minutes of each other for ALEXIS to label them as a flare. The pipeline flags the mean date-time of the three peaks as a tentative flare peak time. This is accomplished by converting date-times into UNIX time stamps, normalizing to the largest UNIX value, sorting the time stamps, and aggregating entries within a 2 minute window. This time stamp is tentative because ALEXIS will later define the solar flare X-ray magnitude using the 2 s XRS data.

To define a flare's X-ray magnitude, ALEXIS queries the GOES XRS availability SQLite database, which has the 2 s long- and short-band flux. ALEXIS then removes the scaling factor for XRS-B in order to keep the flare X-ray magnitude assigned by ALEXIS calibrated to the next generation of GOES instruments (see the GOES XRS read-me file).[3] A peak-finding algorithm is run on the XRS 2 s cadence where the largest peak is detected in a 2 minute time window of the tentative time stamp. The resulting XRS flux value and time stamp pair are now the final X-ray label and X-ray peak time.

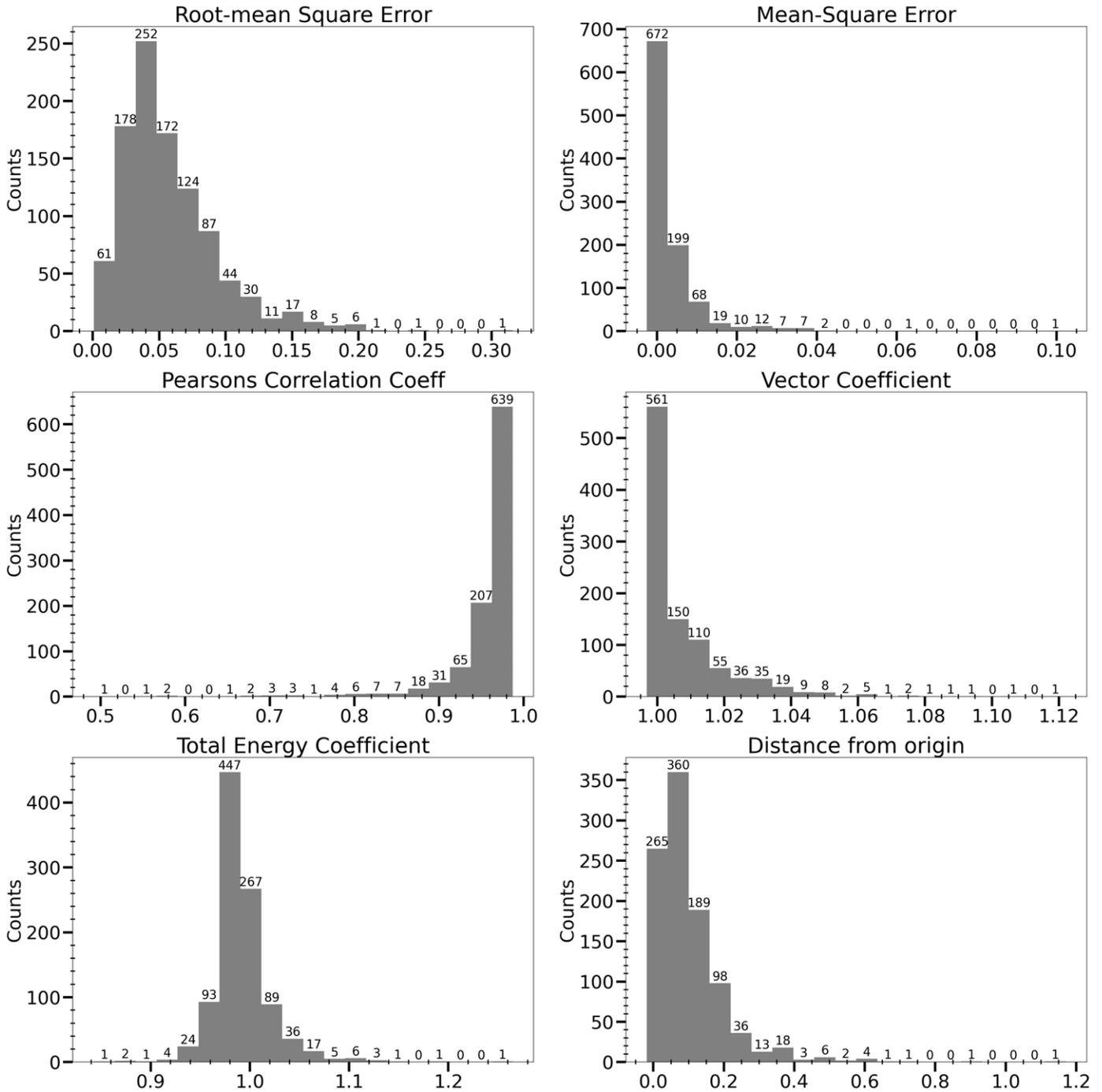

**Figure 8.** Distribution of linear evaluation metrics. Each tile shows the linear agreement between the 1000 hr time ranges ALEXIS has optimized for. Top left: RMSE (Equation (3)). Bins have boundaries between 0.008 and 0.338 in 20 increments of 0.15. Top right: MSE (Equation (4)). Data have been binned at boundaries $7.537 \times 10^{-5}$ and 0.103 in 20 increments of 0.005. Middle left: Pearson's correlation coefficient (Equation (2)). Data have been binned at boundaries 0.559 and 1.02 in 20 increments of 0.0219. Middle right: vector similarity scalar (Equation (6)). Data have been binned at boundaries 1.00 and 1.11 in 20 increments of 0.005. Bottom left: total energy (Equation (5)). Data have been binned at boundaries 0.758 and 1.34 in 20 increments of 0.029. Bottom right: distance from origin (Equation (7)). Data have been binned at boundaries 0.011 and 1.18 in 20 increments of 0.059.

ALEXIS has now defined the location, magnitude, and the HMI AR in which a flare occurred.

## 5. Results

We summarize the results of the ALEXIS pipeline. The agreement between the XRS data and the candidate regions found in the EUV/soft X-ray image time series is discussed in Section 5.1. The accuracy and specificity of the flares reported by canonical databases that have occurred within our 1000 hr time ranges are defined and described in Section 5.2. Flares that were previously unidentified are also discussed in Section 5.2.

### 5.1. Sample Distribution Linear Optimizer and Evaluation Metrics

ALEXIS minimizes the difference between the XRS time series and a time series of discrete regions on the solar surface from multipixel images. We evaluated the distribution of the fit that best reconstructs the single-pixel X-ray ground truth reconstructed by the multipixel components in the EUV or soft X-rays (see Section 4.5).

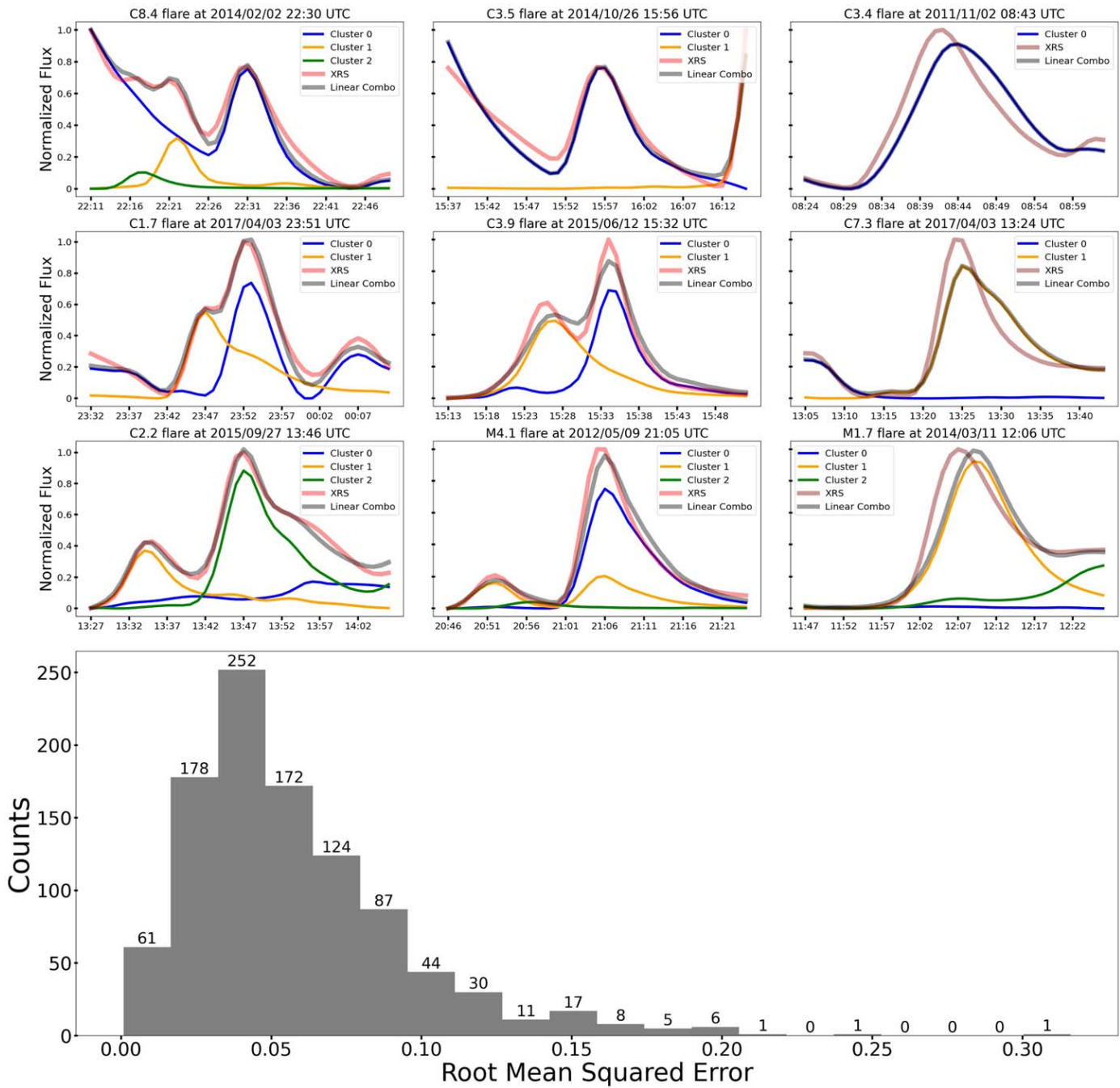

**Figure 9.** Example of predicted signal vs. XRS signal for different RMSE bins. We show the fit of multiple or single contributions to the XRS signal. The XRS signal is in maroon, with the linear combination in gray, and each of the components on the linear combination in their respective colors (cluster 0 (blue), cluster 1 (orange), cluster 2 (green)). The *x*-axis is date-time with a tick mark every 5 minutes. The *y*-axis is normalized flux. Left: $0.0 \leqslant \text{RMSE} < 0.05$; center: $0.05 \leqslant \text{RMSE} < 0.1$; right: $0.1 \leqslant \text{RMSE}$: Bottom: enlarged example of RMSE distribution also seen in Figure 8.

Overwhelmingly, the combination of X-ray signal as recorded by GOES15-B and the multipixel AIA-131 Å image produces the best fit (47% of the sample). This result does not defy logic: AIA-131 Å tracks flaring material and flaring regions associated with regions capable of reaching temperatures in the $10^7$ K characteristic of Fe XX and Fe XXIII (see

Table 1). In addition, GOES15 was operational for most of the time range from 2010 to 2020, while the XRS instrument on board GOES13 started collecting data in an operational capacity from 2015 onward. The next most common combination is GOES13-B with AIA-131 Å (13%), followed by GOES15-A with AIA-131 Å (10%), and finally GOES15-B with AIA-193 Å (8%). The remaining 22% of samples are distributed among the other EUV X-ray mappings. A simplified plot showing the distribution of the best-fit metadata can be found in Figure 7.

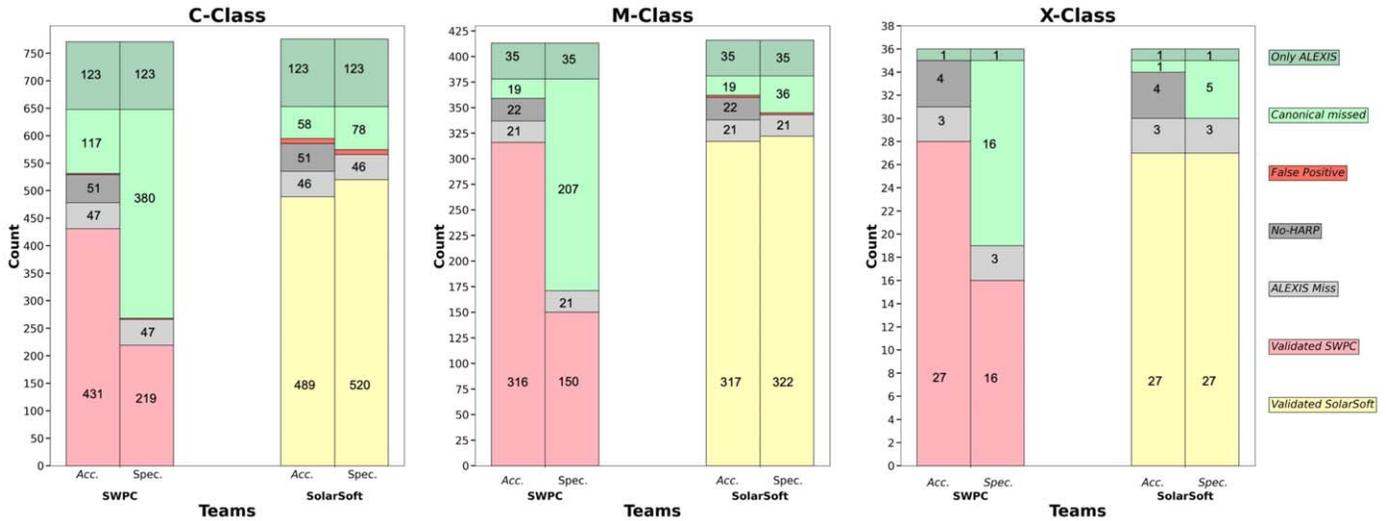

**Figure 10.** ALEXIS catalog vs. SolarSoft vs. SWPC: bar plot showing how the flare subsample evaluated in this proof of concept compares to the entries found by ALEXIS. Each column in the plot is separated by flare magnitude bins (left: C-class flares; center: M-class flares; right: X-class flares). For each team (SWPC and SolarSoft), we show the census results in terms of accuracy (labeled Acc.) and specificity (labeled Spec.). From the bottom up: light pink follows entries in the SWPC that are correct, while yellow shows entries in SolarSoft that are correct. Next is the "gray zone," where we show in light gray events that ALEXIS missed, and in the darker gray events ALEXIS found but that do not have a HARP. If the canonical teams have false positives, they are shown next in dark red. The "green" zone follows, with mint green showing events that the SWPC or SolarSoft incorrectly cataloged. The darker green on top shows events that are unique to ALEXIS, representing new flares that have never been described before.

Figure 7 also shows the distribution of the regularizing value needed to create the best fit. Minimal regularization is needed for approximately 36% of the sample. By creating all possible combinations and exploring the full parameter space, we minimize the need for regularization, but this does not mean that a simple least-squares fit will always create the best convex fit. For example, let us imagine that we are returned three candidate regions (see Section 4.2 for a definition of the candidate region). Assume that only two regions are necessary. The convex fit with the three regions will require more regularization to fit the data than the other training matrix with information from the two necessary regions only. The peak at lower regularizing values indicates that many time series may not need or benefit from strong regularization because those models are relatively simple, with few or no candidate regions being penalized or discarded. For the other two-thirds of the sample, the distribution suggests that most of the models lean toward moderate to strong regularization (higher lambda values), which is often a strategy to avoid overfitting in LASSO regression. The ALEXIS pipeline has a general preference toward models that are more regularized, possibly to reduce multiple candidate regions or improve generalization by simplifying the model. Regardless, we believe that the optimized LASSO is an interpretable, simple, and computationally inexpensive tool for our needs.

The distribution of the evaluation metrics for the best fit is shown in Figure 8. Recall that a perfect fit between the X-ray and linear combination vectors will yield metrics RMSE = 0, MSE = 0, Pearson's = 1, total energy = 1, and vector similarity = 1. MSE is the average squared residual that penalizes large differences between the predicted value and the ground truth. Eighty-six percent of the sample lies within an MSE < 0.01, suggesting that there is small variance between the X-rays and the linear combination vector. RMSE provides insight into the average magnitude of the residuals between the ground truth and the predicted vector and can be interpreted similarly to the standard deviation between the data sets. The RMSE distribution peaks at RMSE values of 0.05 with tails that expand into the worst RMSE values of 0.30. Nevertheless, interpreting the RMSE and MSE depends on the dynamic range of values that the data are allowed to have; ALEXIS normalizes all data to be between 0 and 1. We show examples of convex optimized signals from the sample presented in this paper representative of different bins in the RMSE space in Figure 9 for greater interpretability.

The total energy in the system ($E_{tot}$) is a coefficient that scales the energy of the linear combination to the energy of the X-ray signal. We describe the distribution of total energy in the system as optimal with a peak at $E_{tot}$ at the nominal value of 1.0 with tails that extend both to lower and higher coefficients. If the value of this coefficient is greater than 1, we have underestimated the energy. If the value of this coefficient is less than 1, we have overestimated the energy. ALEXIS over-estimates the energy in 55% of the samples and underestimates it for 45%. The other optimized metric, a weight that scales the two vectors or the vector similarity coefficient, is also at optimal values of $V_{full} = 1$. With an interpretation similar to the $E_{tot}$ coefficient, ALEXIS marginally underestimates the X-ray signal. The Pearson correlation coefficient also measures the linear relationship between the vectors. The strength and direction of the linear relationship between variables span values between −1 and 1. The sample in this proof of concept follows optimal values for the Pearson function, indicating a good agreement between the trend between the predicted and ground-truth X-ray values.

Taken together, the ALEXIS pipeline has the capacity to reproduce the linear relationship between the full-disk X-ray flux and discrete regions from the full solar disk to a good degree of agreement. Our results imply that the majority of the X-rays in the nominal field of view produced in the 0.5–8 Å (1.5–25 keV) band can be attributed to a sample of discrete regions on the solar surface.

### 5.2. ALEXIS versus SolarSoft versus the SWPC

We compare the results from ALEXIS with the results published on the HEK database for C-class flares and above from 2010 March to 2020 May as reported by SolarSoft and the SWPC (Figure 10). The main goal of the ALEXIS pipeline is to be able to identify the exact location of solar flares in order to map an event to the HARP and the magnetic field features associated with them. We compile how the metadata presented by SolarSoft or the SWPC will map to a HARP under two key metrics: accuracy and specificity. Note that, regardless of technique (accuracy or specificity), multiple HARPs may map to the same flare. This problem will arise because HARPs can spatially encompass multiple NOAA ARs and also multiple HARPs can be associated with the same NOAA AR. Overlapping HARPs are a general problem not unique to ALEXIS; disambiguation of the HARPs when they cannot be spatially separated is an outstanding problem in the field. It is also worth noting that the flare magnitudes reported by ALEXIS have been scaled to the next generation of GOES telescopes (16, 17, and 18). This scaling requires removing the correction factor, which will increase the flare magnitude of all refactored flares (see the GOES 16 read-me file)[3] for details about scaling factors applied to new and historical flare magnitudes). Below, we describe the resulting census reported by ALEXIS.

*Accuracy.* Any researcher wanting to identify the HARPs that have an associated flare must approach this question in two steps. The first step is to try to map the NOAA AR of a flare to the HARP number using the NOAA_ARS and NOAA_AR keywords; this is common practice in the literature. NOAA ARS identify all NOAA ARs related to the HARP during the ≈14 days transit period across the Earth-facing solar disk. The NOAA_AR identifies the best match for a HARP from the group of ARs in the NOAA_ARS keyword metadata. If no match is found, one proceeds to map from the location metadata reported by ALEXIS to the spatial span of a HARP. We call this method "accuracy"; teams default to mapping from the AR to the HARP and then bypass to the spatial location if the NOAA AR number is missing. With the accuracy method, scientists do not know exactly where within the HARP a flare occurred but can attribute it to the general spatial span of the NOAA AR or HARP. On the other hand, we also evaluate the specificity of the canonical entries.

Specificity requires canonical entries to lie within ALEXIS's integration boxes. This will enable geometric features from AIA/HMI data to be generated and applied to the prediction of rare events (E. Jonas et al. 2018; Y. Chen et al. 2019). These geometric features are calculated by embedding the pixel representation of the exact location of the flare into a scalar. This scalar value encodes spatial information on the activity of the flare. Specificity allows embedding of pixels that are solely related to the flare, excluding information that is not relevant to the phenomena.

*No flares returned by ALEXIS for the 1 hr time range.* First, it is important to note that ALEXIS fails in several instances; the pipeline does not return a flare peak time for 4% or 46 one-hour time ranges where ALEXIS reports no flares have occurred. These include three X-class flares, 16 M-class flares, and 27 C-class flares. From this sample that ALEXIS does not recover, 35% are linear combinations that use the SXI images so that the slow temporal cadence of the signal causes peaks to be missed. More than half (60%) of the missed sample can be recovered by increasing the flare detection time window to three data points or 3 minutes (see Section 4.6). Upon visual inspection, 93% of the emission from the flare locations have characteristic increase and decrease in emission and would be interpreted as the correct flare location. Two out of three missed X-class flares exceeds the hour of data under analysis; no peak is found because we need more data.

*ALEXIS false positives and missed canonical flares.* Of the 1211 flares reported by ALEXIS, 5% are false positives. These false positives include 49 C-class flares, 10 M-class flares, and one X-class flare. In total, ALEXIS misses ≈6.6% of the 1061 flares used in this proof-of-concept subsample reported in this paper, leaving 988 flare entries for which we can compare the metadata for all teams. For every entry each team provides, we searched within ±4 minutes centered on their reported flare peak time. If there exists an entry in the ALEXIS catalog within that period, true in time is flagged.

*ALEXIS returns a flare but not a HARP.* There are occasions when ALEXIS will return a flare date-time, magnitude, and location, but is not able to find a HARP. This happens for 77 flares, including 51 C-class, 22 M-class, and four X-class flares, representing 7.8% of the 988 comparable flares explained above. All of these events are flares that are located on the solar border and will struggle with HARP association in the helioprojective plane. ALEXIS entries that do not return a HARP are incomparable when computing the accuracy of a canonical flare entry.

*SolarSoft versus ALEXIS.* Recall that SolarSoft has 1069 events with 11 false positives, leaving 1058 flares, of which ALEXIS misses 70. There are still 988 flares for which ALEXIS and SolarSoft can be compared. Fifteen of these SolarSoft flares lie outside the ALEXIS peak time by ±4 minutes. With respect to specificity, SolarSoft is not within ALEXIS's boundary box for 12%, or 119 of the 988 remaining flares. SolarSoft is not specific for 78 C-class, 36 M-class, and five X-class flares.

On the other hand, the sample we can use to compare accuracy is smaller because we need to exclude the 77 flares for which ALEXIS failed to report a HARP. There are 547 C-class, 336 M-class, and 28 X-class flares, or 911 flares for which we can compare accuracy. SolarSoft is not accurate for 9%, or 78, of the 911 entries (58 C-class, 19 M-class, and one X-class flare). Seventeen flares, or 1.9%, are specific but not accurate (13 C-class, four M-class, and two X-class flares), implying that the metadata have incorrect information for the NOAA AR, but the coordinates reported are correct. Fifty-one flares, or 5.6% (29 C-class, 18 M-class, and four X-class flares), are accurate but not specific, meaning that the NOAA AR has been correctly identified, but the location is beyond the boundaries found by ALEXIS. However, 6.7%, or 61 (45 C-class, 15 M-class, and one X-class) flares, are neither specific nor accurate, but are correct in the timing of the flare peak; the locations for these entries are incorrect in the SolarSoft catalog.

Since accuracy is the technique most used by researchers in the field, we conclude that scientists who use the SolarSoft catalog will need to revisit ≈10% of the 911 comparable flare entries. This includes the 15 entries for which the SolarSoft flare peak time is beyond ALEXIS's peak time by more than ±4 minutes.

Overall, ALEXIS will update the metadata for 15% of the 988 flares in the SolarSoft census, adding metadata for the timing for 15 flares, updating the accuracy and specificity for 61 flares, updating specificity for 58 flares, and updating the accuracy for 17 flares. Of the 1057 events reported in this subsample by SolarSoft, ALEXIS will not be able to tell you anything for 70 (see Figure 10).

*SWPC versus ALEXIS.* Recall that the SWPC has 1061 events with two false positives, leaving 1059 flares, of which ALEXIS misses 71. There are 988 flares for which ALEXIS and the SWPC can be compared (599 C-class, 357 M-class, and 32 X-class flares). The SWPC is not specific for 61%, or 603, of these flares (380 C-class, 207 M-class, and 16 X-class).

On the other hand, in order to compare accuracy, we need to deduct the 77 flares for which ALEXIS returns no HARPS. There are 548 C-class, 335 M-class, and 28 X-class flares, or 911, for which we can compare accuracy. The SWPC is not accurate for 15%, or 136, of the 911 entries (117 C-class and 19 M-class flares). There are three C-class flares that are specific but not accurate, implying that the metadata have incorrect information for the NOAA AR, but the coordinates reported are correct. Forty-four percent, or 396 (216 C-class, 167 M-class, and 13 X-class) flares are accurate but not specific, which means that the NOAA AR has been correctly identified, but the location is beyond the boundaries found by ALEXIS. One hundred and thirty-three (114 C-class and 19 M-class flares) are neither specific nor accurate, but are correct in the timing of the flare peak; the locations for these entries are incorrect in the SWPC catalog. Since the accuracy technique is the most used by researchers in the field, we conclude that scientists who use the SWPC catalog will need to revisit ≈16% of the 911 comparable flare entries, including the six entries that lie outside of ALEXIS's by ±4 minutes peak time.

Overall, ALEXIS will update the metadata of 62% of the 988 flares in the SWPC census, updating the timing for nine flares, updating the accuracy and specificity for 133 flares, updating the specificity for 466 flares, and updating the accuracy for three flares. Of the 1058 events reported in this subsample by the SWPC, ALEXIS will not be able to tell you anything for 70 (see Figure 10).

*Previously unknown solar flares.* In total, ALEXIS identifies 159 previously unknown flares. Events unique to ALEXIS that have not been cataloged before include 123 C-class and 35 M-class flares and one X-class flare. This represents an increase of 15% from the original sample of ≈1057 flares ALEXIS sought to find. It is important to note that the simple techniques ALEXIS employs have still not identified new flares such as those in Figures 15, 16, and 17. Manual human verification is still necessary to have a complete census of solar flares as observed by the X-ray and EUV convex optimized signals.

## 6. Discussion

ALEXIS uses contemporary computing techniques to maximize the scientific return of the AIA, the SXI, and the XRS for solar flare science. We have shown that ALEXIS can recreate the X-ray emission integrated across the whole solar disk using the emission from a weighted linear combination of discrete regions on the solar surface, leveraging multiwavelength and multi-instrumental data. In particular, ALEXIS metadata preprocessing allows SXI data to be incorporated into the pipeline, which opens up 10 yr of rarely used archived data for use in detailed solar flare studies using the tools created by The SunPy Community et al. (2020). Additionally, since AIA data files are large, most flare prediction and flare localization studies down-sample the data in both the spatial and temporal domains, and rely on limited external data products, with simplifying assumptions (e.g., R. A. Angryk et al. 2020; S. K. Rotti et al. 2020; K. van der Sande et al. 2022). ALEXIS depends on neither; it starts from the raw full cadence and resolution data stream and returns a processed data product independent of any external biases, assumptions, and conclusions. In the following, we discuss examples of the science that can be enabled by our framework.

Differential analysis of XRS data can trigger solar flare detection: ALEXIS needs only a date-time and a time range in order to independently download, parse, and produce the locations, timing, and magnitude of solar flares, beginning with the raw data available. The date-times passed onto the pipeline can be created by applying a differential analysis of the resampled X-ray signal gathered by integrating across the full solar disk. In contrast to historical flare detection from the single-pixel XRS instrument, ALEXIS allows the slope of the underlying time series to flag time stamps when flares might have occurred. Allowing the underlying data to identify the time stamps when a flare peak occurred has several benefits. For example, differential analysis allows for the identification of multiple flares within the same time range, which would not be possible under the definition the SWPC uses to flag the time at which a flare occurred. In contrast, other teams consider the same XRS signals as a single event.

The same XRS signal can be composed by emission from multiple locations on the Sun. The pipeline has also shown that a single XRS signal can be composed of multiple flaring regions located a significant distance away from each other. It is validated against the ground-truth XRS flux; sometimes, without the emission from multiple regions, one cannot recreate the X-ray data as integrated across the full disk. From this proof-of-concept sample (see Section 3.3), there are four ways in which more than one region is needed to fit the XRS flux. First, the coupling could be just noise; the weight returned by LASSO could not differentiate the need for that region. Second, more than two regions have a drastic increase in emission within minutes of each other. Synchronous flares can also happen where more than two regions flare at the same point in time. Finally, ALEXIS can also recover the background emission before or after the flare. This last result provides motivation for expanding ALEXIS into near-real time. ALEXIS can provide regions of interest for 1 hr before the current observation window. Those regions can be tracked in time, providing priors as data are being collected. When a whole new hour of data has been collected, the pipeline can be run again to see if there are new regions that are of interest.

ALEXIS creates an event list that pinpoints the location of solar flares. Flare prediction would greatly benefit from identifying the exact location of instabilities that lead to magnetic field reconnection. The pipeline identifies regions on the solar surface that are bright for an extended period of time persistent across multiple wavelengths zoomed in to a minimum of 50″. ALEXIS updates the metadata for the flares as compared to canonical databases and finds more (see Section 5.2). The event list presented in this paper is not error free, must be scrutinized by the user, and should be interpreted as a supplement to be used together with other flare catalogs. Knowing the exact location of a large flare census can help scientists uncover statistical relationships in observations and simulations.

For example, can we encode the particular set of circumstances that make an AR erupt? By locating the specific boundary within an AR where magnetic reconnection occurs, ALEXIS can help encode the local parameters of a flare. The ALEXIS localization can help to better detect and characterize the chirality of the magnetic field measured by the HMI. Scientists can also include EUV images to encode the orientation of sigmoids. Combining HMI and EUV images, we can now better identify the exact region where the polarity inversion lines occur, which leads to identifying the exact footprint of the flux rope. Together, this can help determine the orientation of the eruptive loop (right- or left-handed loop). EUV signatures such as the sigmoidal structures (S or reverse S) and the posteruptive arcades (skewness of EUV loops and polarity of the underlying magnetic field) can now be determined with greater fidelity. Knowing the orientation and location of the flux ropes in advance can help provide information about the impact of a CME on Earth.

*Flare duration, solar flare magnitudes, and revisiting solar flare phases.* ALEXIS has shown that multiple flares can compose the same XRS signal, proving that the current method of using the single-pixel XRS-B flux to characterize solar flares is insufficient to fully describe the system. For decades, scientists have tried to distinguish different phases from solar flares by evaluating the timescales of solar emission across the different parts of the electromagnetic spectrum, which is generally studied and identified using X-ray signatures (H. S. Hudson 2011; A. O. Benz 2016). The general consensus is that most flares exhibit phases such as (1) the preflare phase, (2) the impulsive phase, and (3) the decay phase. A new technique using ALEXIS's signal decomposition will allow us to better understand and characterize the phases of the energy release related to solar flares. We suggest evaluating a new way to use the EUV data for the purposes of solar flare start, peak (flare magnitude), and stop times, and consequently the flare phases must be reevaluated.

*Sympathetic and synchronous flares.* Solar eruptions happening at different locations of the Sun at the same moment because of some physical connection are defined as sympathetic flares. In other words, if one flare occurs in one location, how does that event increase or influence the probability of another region erupting? The relationships between these events are important to understand because they provide local and global information on the mechanisms that govern the initiation and propagation of solar flares. ALEXIS has shown that flares can occur from different ARs at exactly the same time (Figures 22, 23, and 24). ALEXIS will help answer questions that could not be posed using canonical catalogs (R. Mawad & X. Moussas 2022, and references therein). For example, what is the frequency of more than one AR erupting simultaneously? Is there a physical connection? To what extent did the increase in emission of a nonflaring region increase the probability of emission in a flaring region? What are the drivers of large flares?

*Increasing the solar flare catalog.* The number of solar flares is used as a proxy for different types of helioresearch, and ALEXIS's new census will help refine these types of studies. For example, the class-imbalance problem in the ML space attempts to predict an event that rarely occurs; the amount of data ingested for flare prediction tasks are qualitative of a quiet, nonerupting star. ALEXIS may help mediate this event by increasing the number of large flares and small flares that have occurred. Also, magnetic reconnection and how this phenomenon begins, propagates, and influences plasma dynamics has connections and implications for both astrophysical and laboratory phenomena. The larger census that ALEXIS provides will help bridge the gap between observation and simulation from our closest extreme laboratory that is our Sun and how that knowledge can affect modeling of stellar, black hole, and accretion physics, as well as plasma physics and energy production on Earth (M. Hesse & P. A. Cassak 2020; D. I. Pontin & E. R. Priest 2022).

*Assigning correct labels.* Further, our work has shown that 15% and 62% of flares from canonical solar flare catalogs are incorrectly localized, which increases confusion in ML models. In other words, using canonical locations will map to the wrong AR undergoing magnetic reconnection, which provides improper input to flare models. On the other hand, ALEXIS demonstrates that it can correctly map multiple flares to multiple ARs within a high-cadence data stream. Previous work more coarsely mapped ARs, both temporally and over the surface of the Sun, often missing multiple flares and, therefore, incorrectly assigning magnetic field metadata of several flares as one temporal event. A more complete census of solar flare events provided by the ALEXIS catalog will help mitigate this issue.

*Augmenting ML features for solar flare prediction.* The output of ML models is only as good as the data that are fed into these statistical models. Several metrics exist to evaluate the goodness of fit, such as the true skill score (TSS). The TSS quantifies how well a binary classification model distinguishes between positive and negative cases. In the literature, regardless of the amount of data or complexity of the model used, the same mean evaluation metrics are returned (TSS $\sim 0.7 \pm .13$). The ALEXIS catalog will be able to augment the four main groups of the time-series features used for solar flare prediction. The first set of features, the magnetic field properties associated with enhanced flare production (K. D. Leka & G. Barnes 2003) calculated and presented as SHARPs (M. G. Bobra et al. 2014), will be augmented by refitting magnetic parameters to the exact vicinity of the flaring location. The second set of features, an exponentially decaying function that encodes the flaring history of the AR (E. Jonas et al. 2018), will be augmented by providing an updated census of how many flares have occurred for each AR. The third set of features, visual information from images using computer vision, will be augmented by ALEXIS's spatiotemporal specificity. This will allow computer vision architecture to encode only data related to a flaring event, discarding visual noise from pixels not directly related to the flare. Until now, predictive models have attributed flares to the general location of an AR, which poses problems when using computer vision techniques that are sensitive to the distance between interesting pixel values. The fourth feature, a running tally counting the number of flares that have occurred in the AR over time (H. Liu et al. 2019), will be updated.

*Broader impact.* There are several interested parties whose work depends on accurate depictions of eruptive events and will be affected by this work. For example, the NCEI and the SWPC will benefit from our pipelines that use the AIA to verify, validate, and update their data products. ALEXIS's updated census will help refine analysis that attempts to understand coronal heating using the GOES catalog (J. A. Klimchuk 2006; J. P. Mason et al. 2023) or predict the remaining duration of a flare (J. W. Reep & W. T. Barnes 2021). In addition, modeling with hydrodynamical codes will need constraints from our work (J. W. Reep & V. S. Airapetian 2023). The USAF and private industry (T.-W. Fang et al. 2022) will depend on our flare models and data products. Heliorelated cubesat groups will better understand what the best launch time windows are in order to maximize their scientific return. Also, historical analysis and publications (such as D. F. Ryan et al. 2012; C. J. Schrijver & P. A. Higgins 2015; Y. Fu & B. T. Welsch 2016; M. Jin et al. 2016; J. W. Reep & K. J. Knizhnik 2019) can be reexamined. Future Artemis missions, Moon bases, private and government space endeavors, and national security infrastructure will all benefit from better space weather prediction tools.

### 6.1. Limitations of ALEXIS

We identified some limitations of the ALEXIS pipeline and outline possible improvements. In general, ALEXIS has difficulty dealing with HARP and NOAA AR associations on the solar limb. The HARP boundary boxes can overlap, causing the spatiotemporal technique to assign multiple HARP numbers to the same event. Both of these problems are not unique to ALEXIS and are nontrivial to solve. This problem arises in data products such as the NOAA event list, the HARP magnetic field patches, and the SHARP space weather magnetic field features.

One potential avenue to address is the optimization of the hyperparameters *eps* and the member number in DBSCAN (see Section 4.2), which can lead to more candidate regions than necessary, increasing our computational load. ALEXIS saves the metadata related to the actual number of cluster members for each of the spatial, temporal, and hyperspectral clustering. Going back into the data to find an optimal range for these hyperparameters could also decrease the number of false positives. However, tuning the hyperparameter on the whole data set will be computationally costly. Scaling the ALEXIS pipeline to Argonne's computer system (J. R. Padial-Doble 2025, in preparation) requires tens of thousands of node hours (millions of CPU hours), since it runs ≈60 times slower than on a 32-core (64 thread) local server, albeit we can run millions of instances of ALEXIS on a single run using hundreds of nodes of the Department of Energy's cluster. The hyperparameters chosen for this proof of concept have proven to be sufficient for our scientific question. They are interpreted to map candidate flaring regions to a location whose emission occurs for a certain amount of time and is consistent across a certain number of wavelengths.

ALEXIS flare detection depends on the X-ray, linear combination, and candidate vector. All have peaks in their signals within ±2 minutes of each other (two data points). ALEXIS fails at finding flares in which the peak of the X-ray and the peak of the EUV/soft X-ray emission differ by more than 2 minutes. Failing to find or finding too many peak times for a time range can be improved in various ways. For example, some of ALEXIS's failures will be resolved with more data; the peak emission has not yet formed. Also, false positives can be decreased by allowing the heuristics of the low-pass filter to be modified to take into account the magnitude of the flare under analysis. This will allow for a more aggressive smoothing for large events compared to smaller flares. Some false positives arise from changes in the signal caused by saturation in the CCD readout and the AEC algorithm (see Section 2.2.2). Wavelet analysis is being considered as a more aggressive approach to identify peak times from a signal.

This iteration of the ALEXIS pipeline does not attempt to define a flare start and stop time. Specifying when the flare started or ended using the single-pixel XRS data will be misleading, since the full-disk X-ray emission is not sufficient or specific enough to describe the phenomena (i.e., a flare may be composed of multiple emission regions). Higher-order differential analysis taking into account the concavity and slope of a function applied to a time-series signal identified by ALEXIS is being tested as a promising, simple, and interpretable solution to identifying the start and stop in the EUVs. Also, there are still flares that the ALEXIS identification pipeline misses (see Figures 15, 16, 17). All of these would be considered new flares but have not been flagged by the pipeline and are not included in Section 5.2. Human verification and a general scientific consensus as to what a flare is in the EUV X-ray convex optimized light curve must be addressed.

### 7. Summary

ALEXIS uses computer vision techniques coupled with clustering algorithms to accurately pinpoint solar flares. The pipeline automatically downloads solar images across a maximum of 11 wavelengths and with a cadence as small as 12 s. For each image, potential flaring locations are identified with peak-finding and density-based clustering algorithms to find discrete regions on the Sun that are continuously bright temporally and hyperspectrally. For each region, ALEXIS calculates the integrated flux centered at those coordinates within a square boundary of different sizes. We used this integrated flux to generate a time series for every wavelength available. This normalized time series is fit with a spline interpolation and resampled at a 1 minute cadence. An offset correction followed by a low-pass filter is used to smooth the data from high-frequency noise. A normalized grid-searched Cartesian product from each integration area is created, and LASSO is used to minimize the difference between both XRS long and short wavelengths and the flux from a linear combination of EUV or X-ray discrete regions. In this way, ALEXIS finds the set of EUV time series from a linear combination of potential flaring regions that are necessary to describe the whole solar surface detected in the X-rays. The best-fit combination of discrete flaring regions is evaluated using five metrics: RMSE, MSE, Pearson's correlation function, the total energy in the system, and the vector similarity. A perfect fit will yield metrics with $RMSE = 0$, $MSE = 0$, Pearson's $= 1$, total energy $= 1$, and vector similarity $= 1$. In practice, the best fit is the linear combination of potential flaring regions with the smallest distance to the perfect fit within the five-dimensional metric plane. The flare magnitude is then defined where there exists a peak in the time series of the X-ray, linear combination, and individual candidate region vectors. Each candidate region is associated in time and space with known HARPS and NOAA ARs. The specificity and accuracy of the flare metadata reported by SolarSoft and the SWPC are compared. After this process is completed for all time ranges of interest, the ALEXIS flare catalog is aggregated and saved for public download.

We summarize our results as follows.

1. ALEXIS is a pipeline that uses the highest cadence and full resolution of the AIA from the SDO and the SXI from the GOES13–15 telescope. It was designed to provide a data-driven pipeline capable of identifying the location and magnitude of solar flares independently of previous data products (SWPC catalog, SolarSoft catalog, HARP database).

2. The pipeline is presented here as a proof of concept to search for 1057 solar flares randomly selected between the period of 2010 May and 2020 March.

3. Solar flares are identified by measuring the agreement between the XRS signal and a linear combination of emission from discrete regions on the solar surface in soft X-rays and EUV.

4. ALEXIS provides an event list and access to the linear fit for all multipixel wavelengths and each XRS band for every operational telescope available. All time series from each 1 hr time range, for every wavelength, will be provided, enabling the study of rapid variability at different parts of the solar atmosphere.

5. ALEXIS returns 1211 flares where 5% are false positives.

6. The magnitudes of the flares reported by ALEXIS have been scaled to the new XRS data as reported by the GOES13–15 read-me file.[3]

7. ALEXIS has 159 new flares that have not been cataloged before. This is a 15% increase to the compared canonical catalogs.

8. ALEXIS misses 7% of the 1057 flare subsample the pipeline was set out to find.

9. If you are interested in accuracy and use the SWPC, you will be correct for 85% of the comparable subsample. If you want accuracy from SolarSoft, you will be correct for 91% of the comparable subsample.

10. If you are interested in specificity and use the SWPC, you will be correct for 39% of the sample. If you want specificity from SolarSoft, you will be correct 88% of the time.

### Acknowledgments

This research was made possible using the public GOES and SDO data created by NASA, NCEI, and the AIA/HMI science teams. This work is supported by the Fisk-Vanderbilt Masters-to-PhD Bridge program and the Establishing Multimessenger astronomy Interdisciplinary Training (EMIT) with NSF 2125764. The authors would like to thank the reviewers and editors for their help and comments. The authors would also like to thank the University of Chicago Computer Science Department for resources and partial funds in the development of this pipeline.

## Appendix A
## GOES SXI

Cleaning SXI files requires filtering for images predefined for solar flare research, finding solar borders using Hough transforms, resizing or padding the image into a 512 × 512 pixel image, and updating the header file to World Coordinate System standard. Different quality flags are related to the failure of any of these steps; a quality flag of 0 will denote images that can be used in the analysis. ALEXIS only uses images from the SXI instrument that have the exposure time and filters optimized for solar flare research, which include TM, PTHK, and PTHN with exposures under 0.2 s. If an image is taken by any other filter, the pipeline will assign a quality flag of 1. A Hough transform is used to find the solar border of the image (Figure 11); if no solar border is found, a quality flag of 2 is applied to the image. Once the solar border is found, the center pixel pair is verified to be within a 20 × 20 pixel boundary box of the solar center reported by the original file. If the center of the Hough transform lies outside this limit, the image is given a quality flag of 3. Also, the SXI instrument suffers from leaking of pixel values due to the nature of CCD instruments. If the image has bad CCD readout, ALEXIS defines its quality flag as 5. For images that are not originally specified as used for solar flare research with the 'FL' flag, we specify the quality flag as 6. If the fits file has no header information, we specify the quality flag as 10. In total, we used 1190 GOES13-PTHK, 1530 GOES13-PTHNA, 70 GOES14-PTHK, 110 GOES14-PTHNA, 336 GOES14-TM, 2180 GOES15-PTHK, 2660 GOES15-PTHNA, and 10632 GOES15-TM. In order to use the SXI, we must update the fits header standard for use with Astropy and Sunpy libraries; a new way to clean and standardize SXI data allows for its inclusion in research-related tasks. To the authors' knowledge, there has not been any science-oriented publications during the lifetime of the SXI detector.

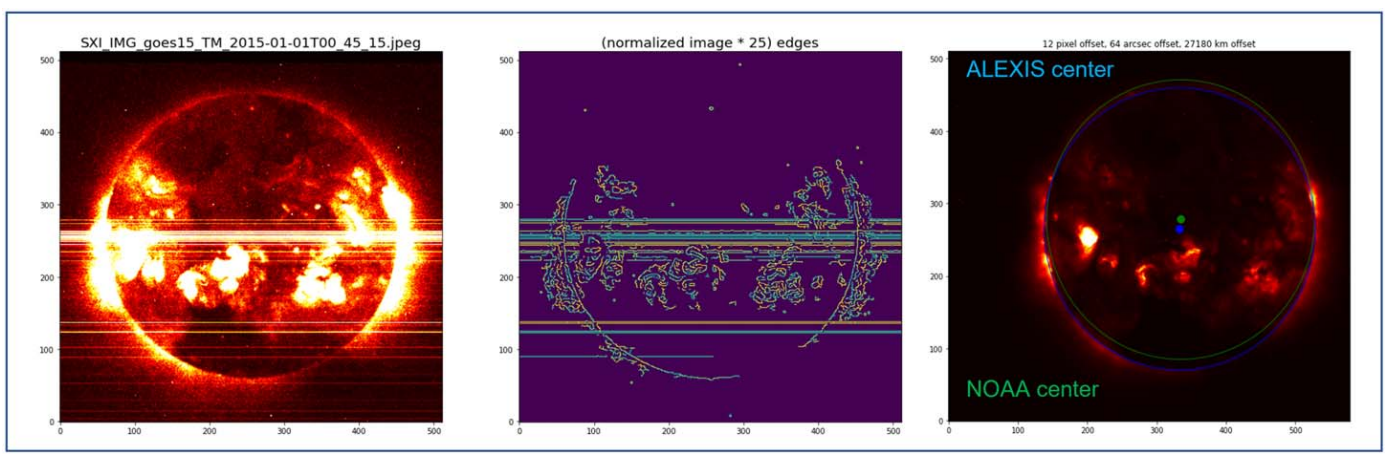

**Figure 11.** SXI cleaning Hough transform, CCD bad readout example, NOAA center vs. ALEXIS. Left: postprocessed SXI image padded to a 512 × 512 image used in ALEXIS. The normalization of the image in this example has been exaggerated to demonstrate CCD bleeding due to the nature of how CCD potential wells behave during readout. Center: edge enhancements using the Canny filter with a $\sigma = 3$ from a Scikit image as a preprocessing step before applying the Hough transforms. Right: Hough transforms are used for identifying simple shapes in images with the help of an edge detector. The circular Hough transform was constrained to identify all pixels that together compose a border within a certain radius determined by the SXI's resolution. On average, unprocessed SXI metadata report the solar center coordinates offset by 64″ (12 pixel space) when compared to the metadata from the ALEXIS processing. Validation of this method was done by reprojecting the SXI image onto AIA pixel space and comparing the location and coordinate conversion bright regions on the surface. The reprojection of images not processed by ALEXIS fail this test.

## Appendix B
## AIA

All of the AIA data downloaded from JSOC are level 1 data products. Processing level 1 data to level 1.5 requires updating the orientation of each image, correcting for the instrument's degradation, and exposure correction. This promotion involves updating the pointing keywords, removing the roll angle, scaling the image to a resolution of $0.''6$ per pixel, and translating the image such that the center of the Sun is located in the center of the image. Pointing keywords to update include CRPIX1, CRPIX2, CDELT1, CDELT2, and CROTA2. The digital number (DN) gain conversion in the metadata is then used to transform the image into units of $e^{-} \ s^{-1}$. Sunpy's AIAPY routines were used to clean AIA data.

Not all images are science quality; images that are science quality are given a QUALITY flag label 0. Nonzero quality flags presented by the postprocessing by the AIA science team are not used. Nevertheless, not all images with a good quality flag were used. Other quality flags identified by the ALEXIS pipeline include images where there are negative value pixels that dominate the image (QUALITY = 4), images where the fits metadata return not a number (NaN) values for the effective area and DN gain (QUALITY = 7), images that fail the registration step because they are not full-disk images (QUALITY = 8), and images that fail the pointing correction because they have NaN values for the telescope metadata that reports the satx, saty, and sat rot. We use 196036 AIA-94, 195279 AIA-131, 191926 AIA-171, 196164 AIA-193, 195680 AIA-211, 195681 AIA-304, and 75415 AIA-335 with good science-quality flags.

## Appendix C
## Compiling Information: Local SQLite Databases

When studying solar flares, one normally begins by creating a sample of events and collecting the metadata related to its intensity, location, and duration. Five databases are created previous to running the ALEXIS pipeline. These SQLite databases contain all known information about the AIA and SXI images used, flares known, HARPs previously detected, and the XRS flux. Information queried from these SQLite databases is returned as pandas data frames and sorted chronologically to a Coordinated Universal Time (UTC) time zone. Databases can be created by the user via scripts found on the GitHub onboarding instructions; remember to change the JSOC user email and the path where you want ALEXIS to run. All databases can be downloaded. Data inserted into these five databases will be filtered to only include information from 2010 May to 2020 March. SQLite databases allow the user to query and insert data in a robust way with Python packages such as SQLalchemy (M. Bayer 2012). The *query_the_data.py* script has several functions for interacting with these databases, and the metadata for each SQLite database can be found in Table 2.

The XRS Availability Database created by the *xray_ftp_query.py* script scrapes, downloads, and creates an SQLite database of the X-ray flux as measured by both XRS band passes (A and B) on board GOES13, GOES14, and GOES15. The XRS sensor value stored is not corrected for the scaling factor described in the GOES XRS read-me file.[3] Nevertheless, when needed, the scaling factor is removed. This is the case when ALEXIS defines the X-ray magnitude of a solar flare (Section 4).

The HMI DRMS Availability Database (*create_drms_hmi_AR_db.py* script) holds the HARP metadata from the Data Record Management System (DRMS) hosted by the JSOC. The boundary box stored here represents the area of the HARP in Heliographic Stonyhurst coordinates taken from the HMI data series *hmi.sharp_720s* at each point in time that it has been observed, not the maximum size of the HARP once it has progressed across the solar surface. The HARPs in this database are not the near-real time data series; HARPs included have already progressed the solar surface and have been preprocessed by the Stanford team. Note that mapping the size of HARPS from HMI scale to AIA scale will require scaling to the plate scale of each instrument; objects in HMI will appear larger than the equivalent area in an AIA image.

The SXI Availability Database (*sxi_availability.db*) holds the URL location of all the SXI images. Note that the URLs for the SXI are no longer visible by browsing, but the FTP links to the SXI images are still active.

The AIA Availability Database (*create_aia_12s_availability.py*) has the URL location of every 12 s cadence image from the AIA.

The HEK Known Flares Database (*create_drms_hmi_AR_db.py*) holds all known AR and solar flares reported by the SWPC and SolarSoft to the HEK portal.

**Table 2**
The Main Index for Each SQLite Database Is UNIX Time Stamps and Can Be Queried via Functions within the *query_the_data.py* Module

| SQLite Database | Columns | Comments |
|---|---|---|
| XRS Availability | Observation date-time | UNIX time stamp (UTC) |
| | Telescope | GOES13, GOES14, GOES15 |
| | Wavelength | A or B |
| | Value | W m$^{-2}$ |
| SXI Availability | Image date-time | UNIX time stamp (UTC) |
| | Telescope | GOES13, GOES14, GOES15 |
| | Download string | SXI image File Transfer Protocol (FTP) location |
| | Data level | Science-quality or raw data flag |
| | File name | Local download path defined by user |
| AIA Availability | Image date-time | UNIX time stamp (UTC) |
| | Exposure time | Seconds |
| | Wavelength | 94, 131, 171, 193, 211, 304, 335 Å |
| | Download URL | JSOC FTP |
| | Data quality flag | Assigned by JSOC |
| | Download file name | User-defined local path |
| HMI HARPs Availability | HARP date-time | UNIX time stamp (UTC) |
| | HARP number | Scalar |
| | HARP boundary box | Heliographic Stonyhurst |
| | Area of boundary box | Microhemisphere |
| | Quality flag | Assigned by JSOC |
| | NOAA_ARS | Comma-separated list of NOAA ARs matching this HARP |
| | NOAA_AR | NOAA AR number that best matches this HARP |
| HEK Known Flares | Start date-time | UNIX time stamp (UTC) |
| | Peak date-time | UNIX time stamp (UTC) |
| | End date-time | UNIX time stamp (UTC) |
| | X-ray class | C, M, X +1.0–9.9 |
| | AR number | Assigned by NOAA |
| | Flare $X$ coordinate | Heliographic Stonyhurst and HPC |
| | Flare $Y$ coordinate | Heliographic Stonyhurst and HPC |
| | Flare bounding box | Heliographic Stonyhurst and HPC |
| | ID team | SolarSoft or SWPC |

**Note.** UNIX time stamps are converted to python date-time objects when queried. Local storage needs are HEK flare db.pickle (SS/SWPC C and above) = 170 MB (24 MB), hmi_drms_AR.db = 612 MB, XRS data (38GB), aia_avail.db (62GB), sxi_avail.db (14GB), SHARP data (418MB). You can find the HMI HARP keywords at http://jsoc.stanford.edu/jsocwiki/HARPDataSeries.

## Appendix D
## Example Zoo

Figures 12–25 show examples of the results from the convex optimization. For a description on how to read these plots, please refer to Figure 3. For a discussion on accuracy and specificity with regards to comparisons between canonical teams and ALEXIS, please refer to Section 5.2.

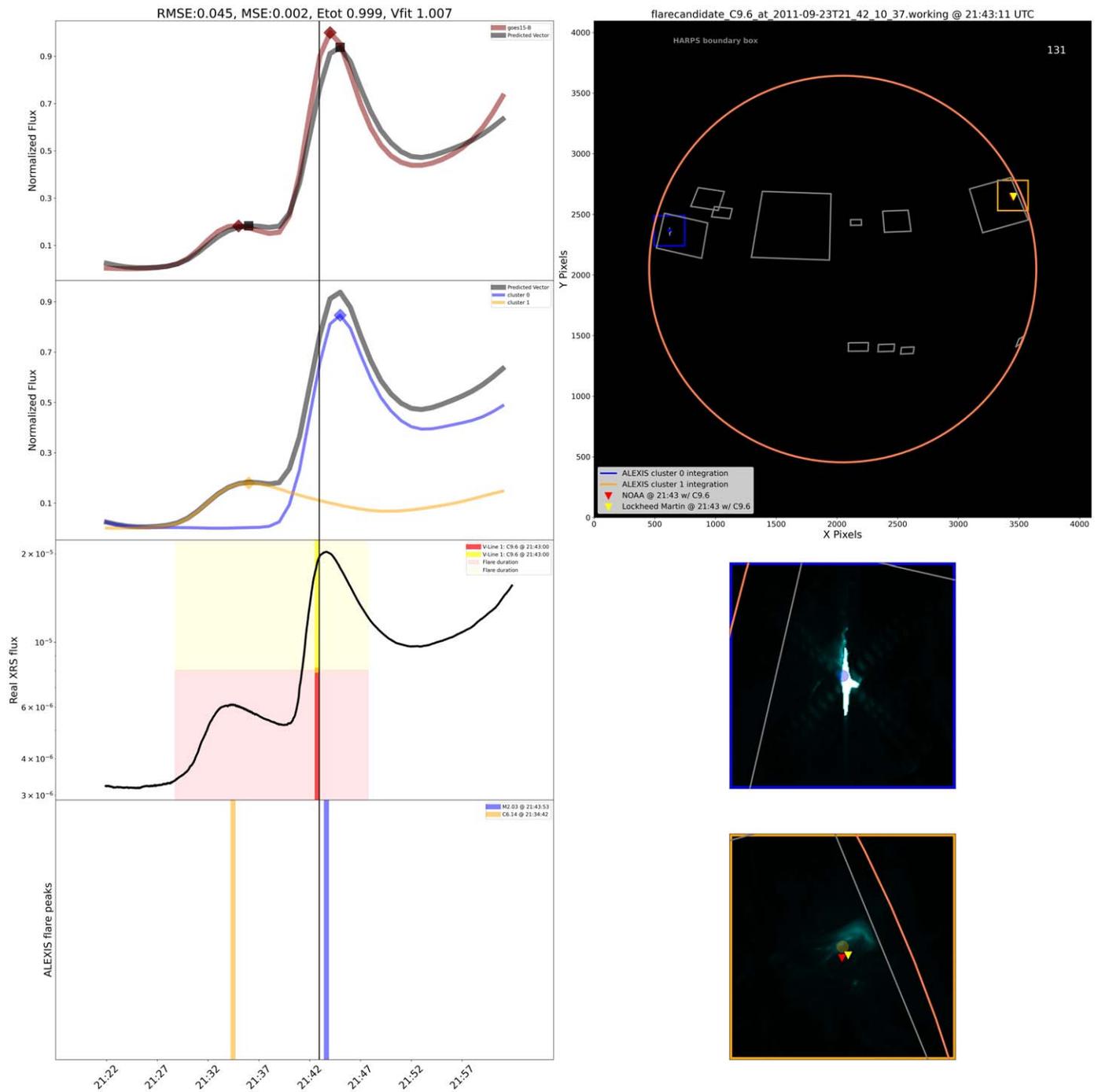

**Figure 12.** Search for C9.6 flare on 2011 September 23 at 21:42:10.37 UTC recreated with GOES15 XRS-B with AIA-131 Å: example where canonical databases wrongly associate an AR to a flare in space and in time. Cluster 1 (orange) is associated with NOAA AR 11295 and HARP 856, and cluster 0 (blue) is associated with NOAA AR 11302 and HARP 892. Both SolarSoft and the SWPC report a flare that occurred within the ALEXIS integration region for cluster 1 (orange) for a C9.6-class flare but occurring at 21:43:00 UTC, which is incorrect. At the time reported by other databases (21:43:00 UTC), ALEXIS locates a flare of X-ray magnitude M2.9 at the opposite side of the Sun's surface at the integration region for cluster 0 (blue). A precursor flare of magnitude C6.14 is detected by ALEXIS occurring at 21:34:42 UTC at region 1 (orange). Note that the duration of the flare the canonical databases mislabelled cannot be completely understood by relying on any method that only takes into account the information from the full solar disk, such as spectral line time series or XRS data. ML models will now include HARP 892 (cluster 0) and HARP 856 (cluster 1) into a positive class; previously, HARP 856 was wrongly placed in the positive class. The SWPC and SolarSoft are not specific or accurate for these entries. Both canonical entries are incorrect in time.

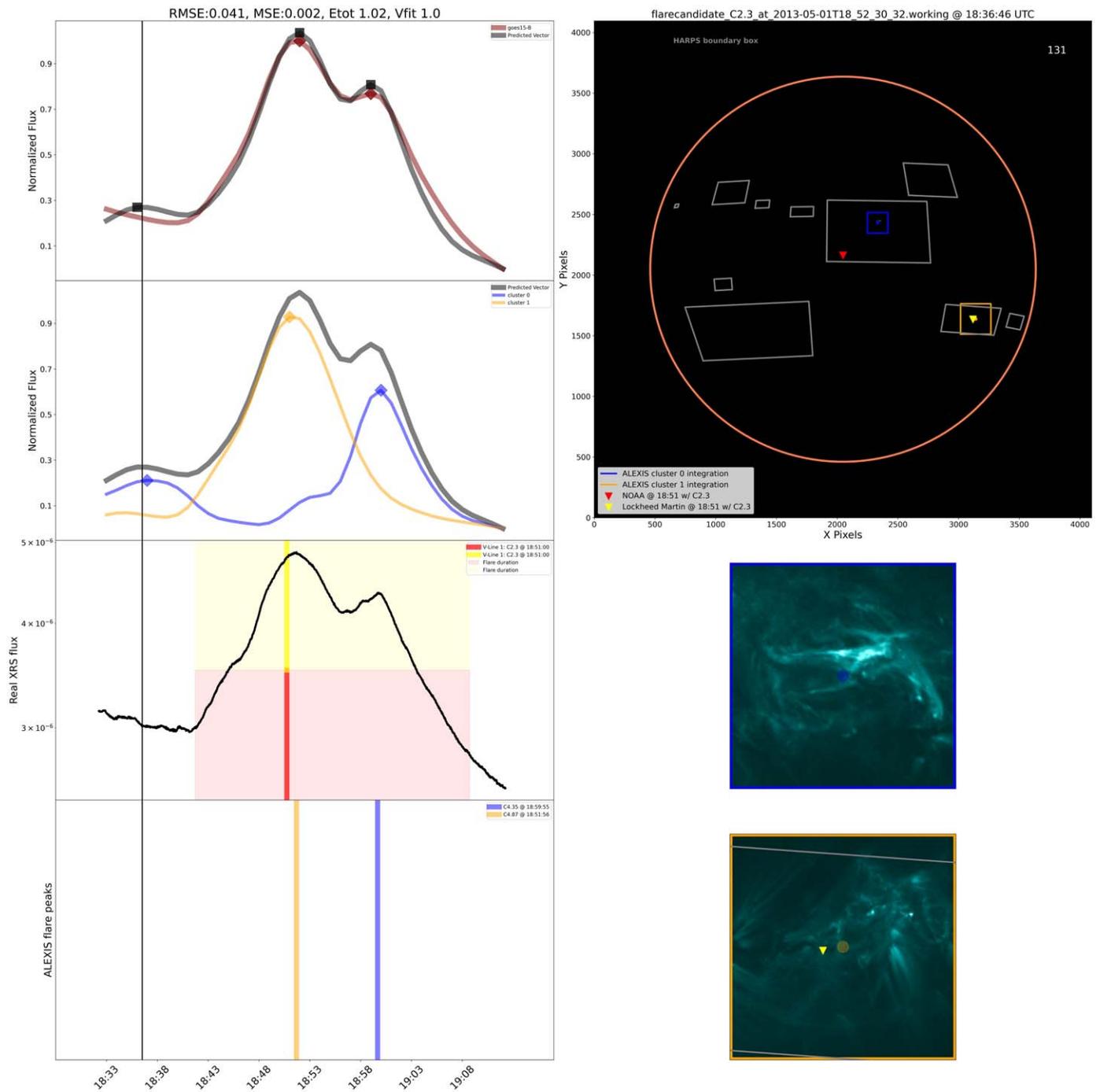

**Figure 13.** Search for C2.3 flare on 2013 May 1 at 18:52:30.32 UTC with GOES15 XRS-B with AIA-131 Å. There are two regions that are necessary in order to compose the single XRS signal. The duration of the flare as reported by other teams is incorrect; the XRS signal could not be reproduced without the C4.35 flare that occurs within the integration region for cluster 0 at 18:59:55 UTC. SolarSoft reported coordinates that are within the integration boundary of ALEXIS's cluster 1 (orange) and thus accurate and specific for HARP 2691 and NOAA AR 11730. SolarSoft correctly identifies a flare occurring at 18:51:00 UTC, but note that the magnitude ALEXIS reports for this flare is C4.87. The signal of the integration region for cluster 0 has an increase to emission prior to the main emission for cluster 1. This raises the question of whether emission from one region may influence the release of energy at another region. Cluster 0 (blue) with HARP 2693 with NOAA AR 11731 was a previously unknown event and will now have the correct label for ML tasks.

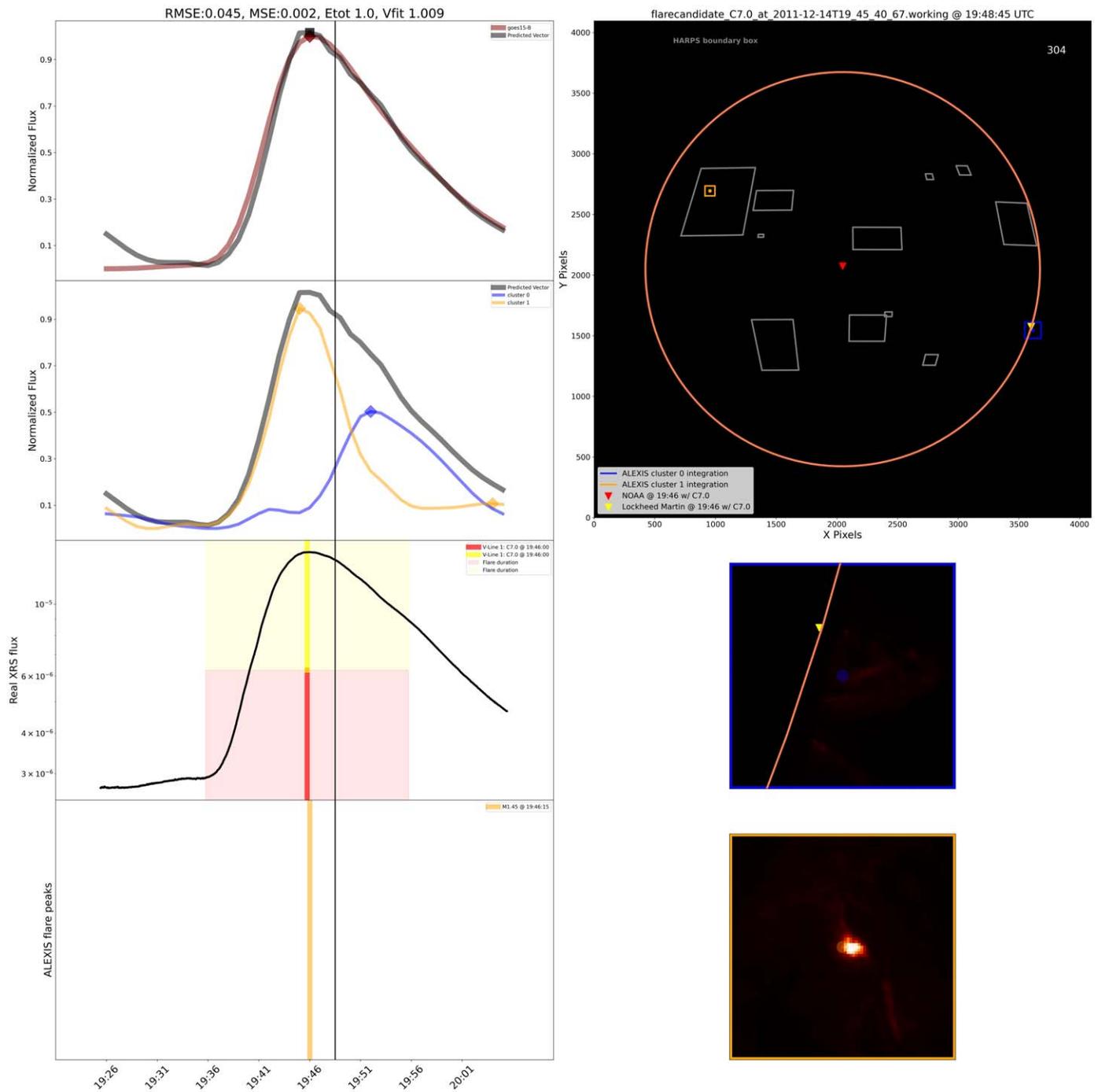

**Figure 14.** Search for C7.0 flare on 2011 December 14 at 19:45:40.67 UTC with GOES15 XRS-B with AIA-304 Å. ALEXIS reports that this flare occurred at 19:46:15 within the integration region of cluster 1 (orange) represented by NOAA AR 11376 and HARP 1183. The emission of cluster 0 (blue) is necessary for the complete recreation of the single-pixel XRS flux. Cluster 0 (blue) could not be associated with any HARP or NOAA AR. This example shows how emission can be hidden and where ALEXIS fails to pick up flare peaks. The SolarSoft and SWPC metadata identify a flare in time but not in space; both the SWPC and SolarSoft are inaccurate and are not specific. HARP 1183 would be placed in the positive class while all other HARPs would be placed in the negative class. Note, this example also demonstrates how the duration of solar flares must be revisited, the need for a multipixel definition for when a solar flare occurred, and the increase in M-class flares number because of the XRS magnitude correction to new GOES instruments.

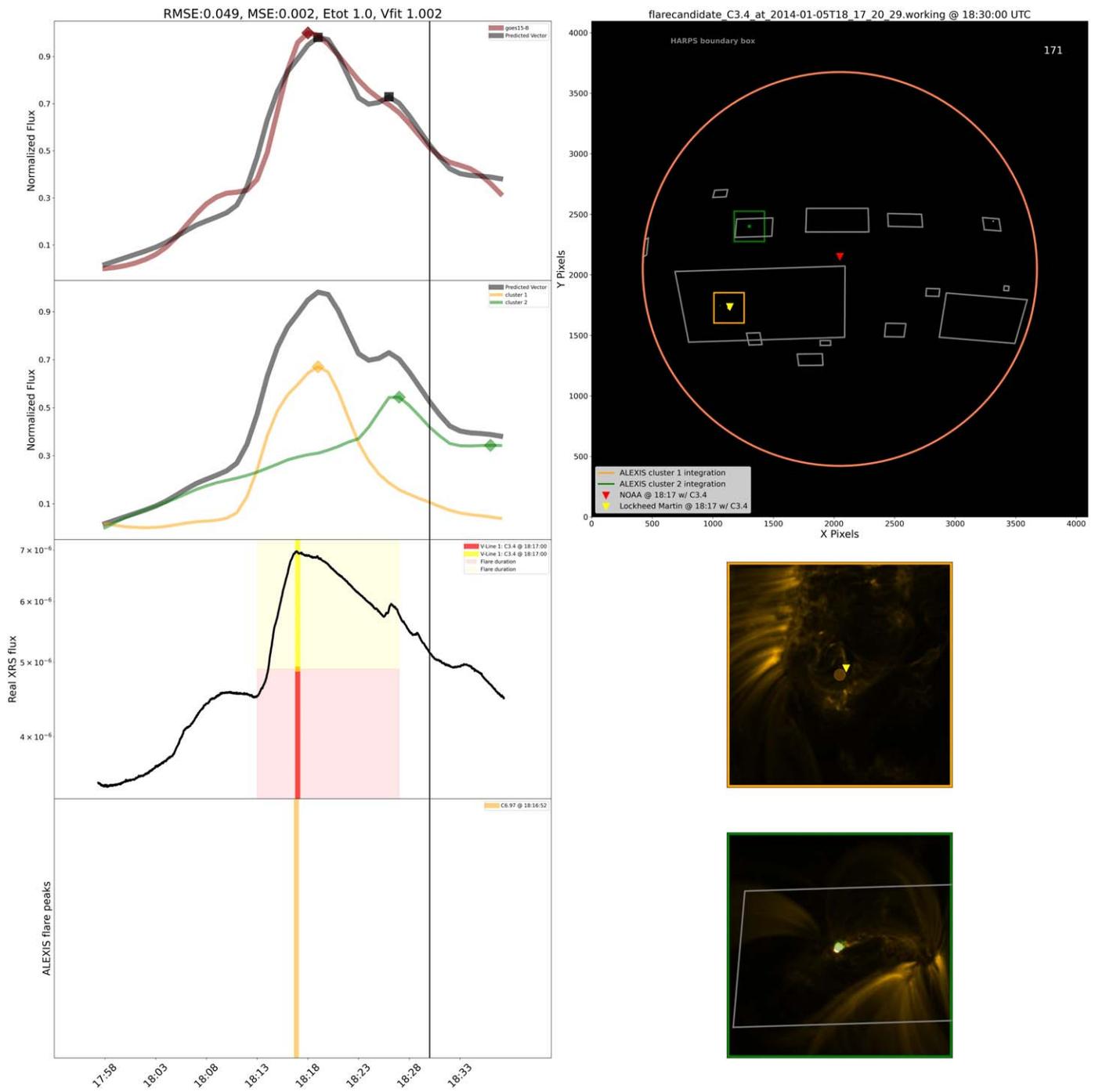

**Figure 15.** Search for C3.4 flare on 2014 January 5 at 18:17:20.29 UTC with GOES15 XRS-B with AIA-171 Å. ALEXIS locates a C6.97 flare at cluster 1 (orange) at 18:16:52 UTC related to NOAA AR 11944 and HARP 3563. Both canonical teams report the same flare related to cluster 1 (orange); SolarSoft returns a C3.4-class flare that is specific and accurate while the SWPC returns a C3.4 flare accurate but not specific. This example shows hidden emission associated with cluster 2 (green-NOAA AR 11946 and HARP 3580) that complicates the classical interpretation of the single-pixel XRS detector. The magnitude of the X-ray class for the flare occurring within cluster 1 will be overestimated because of the contribution to the X-ray signal that cluster 2 provides. Also, ALEXIS fails to pick up the flare at cluster 2 such that the HARP associated with this cluster will be incorrectly labeled when applying ML tasks. This is also an example where scientists might begin to question the influence of emission from one AR toward another.

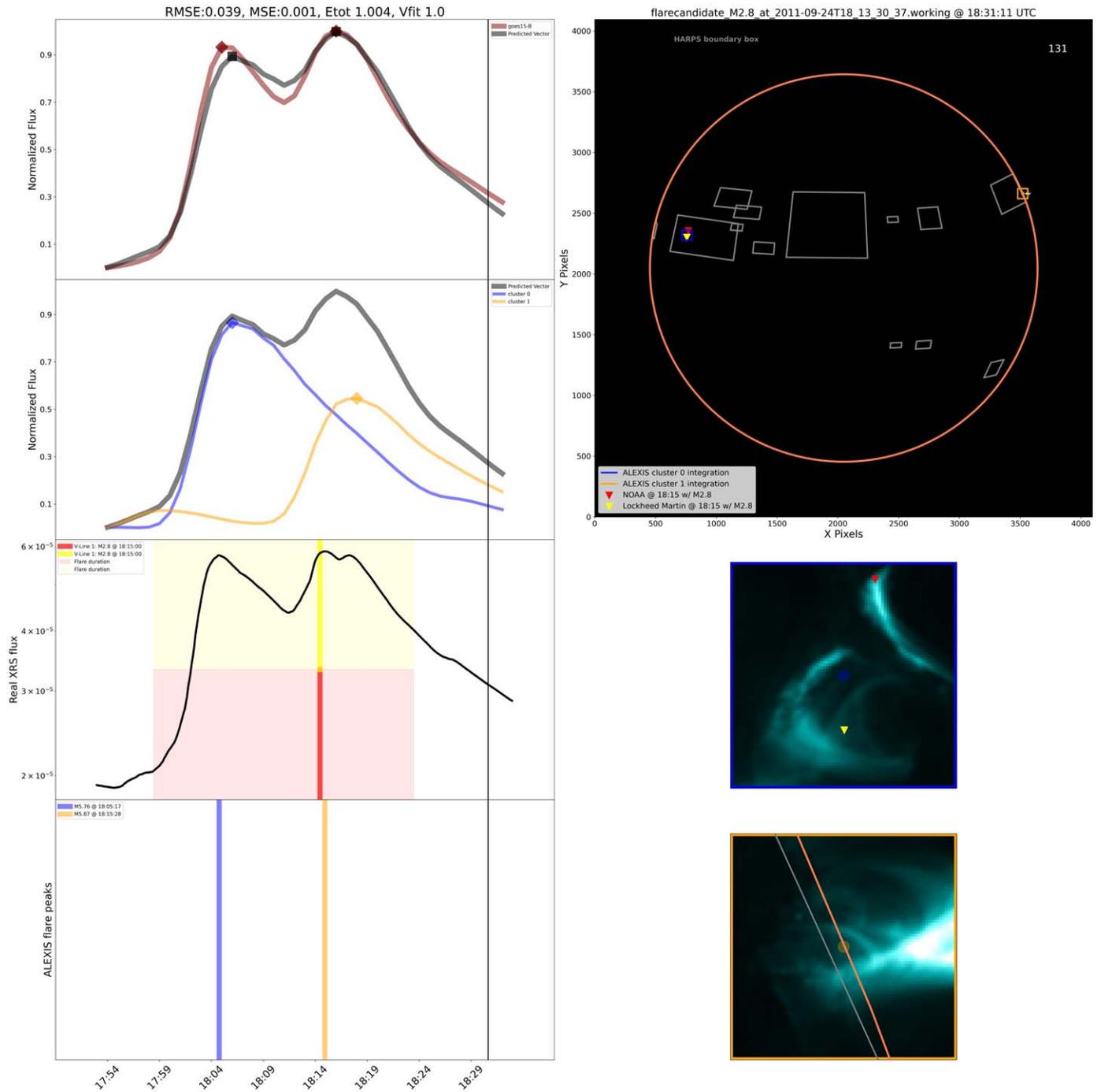

**Figure 16.** Search for M2.8 flare on 2011 September 24 at 18:13:30.37 UTC with GOES15 XRS-B with AIA-131 Å. ALEXIS locates one M5.76 flare at 18:05:17 UTC within the integration box for cluster 0 (blue) and another M5.97 flare at 18:15:28 UTC within the integration box for cluster 1 (orange). SolarSoft and the SWPC located an M2.8 flare at 18:15:00 within cluster 0 (blue) and assigned them correctly to HARP 892 and NOAA AR 11302, but both teams report their flare incorrectly in time. This is an example of the duration of a flare being incorrect and the need for other metrics to label the magnitude of a flare. Without the emission from cluster 0, the magnitude of the flare emitting from cluster 1 (orange) would not be as strong. HARP 856 (NOAA 11295) for cluster 1 and HARP 892 for cluster 0 will be placed correctly into the positive class.

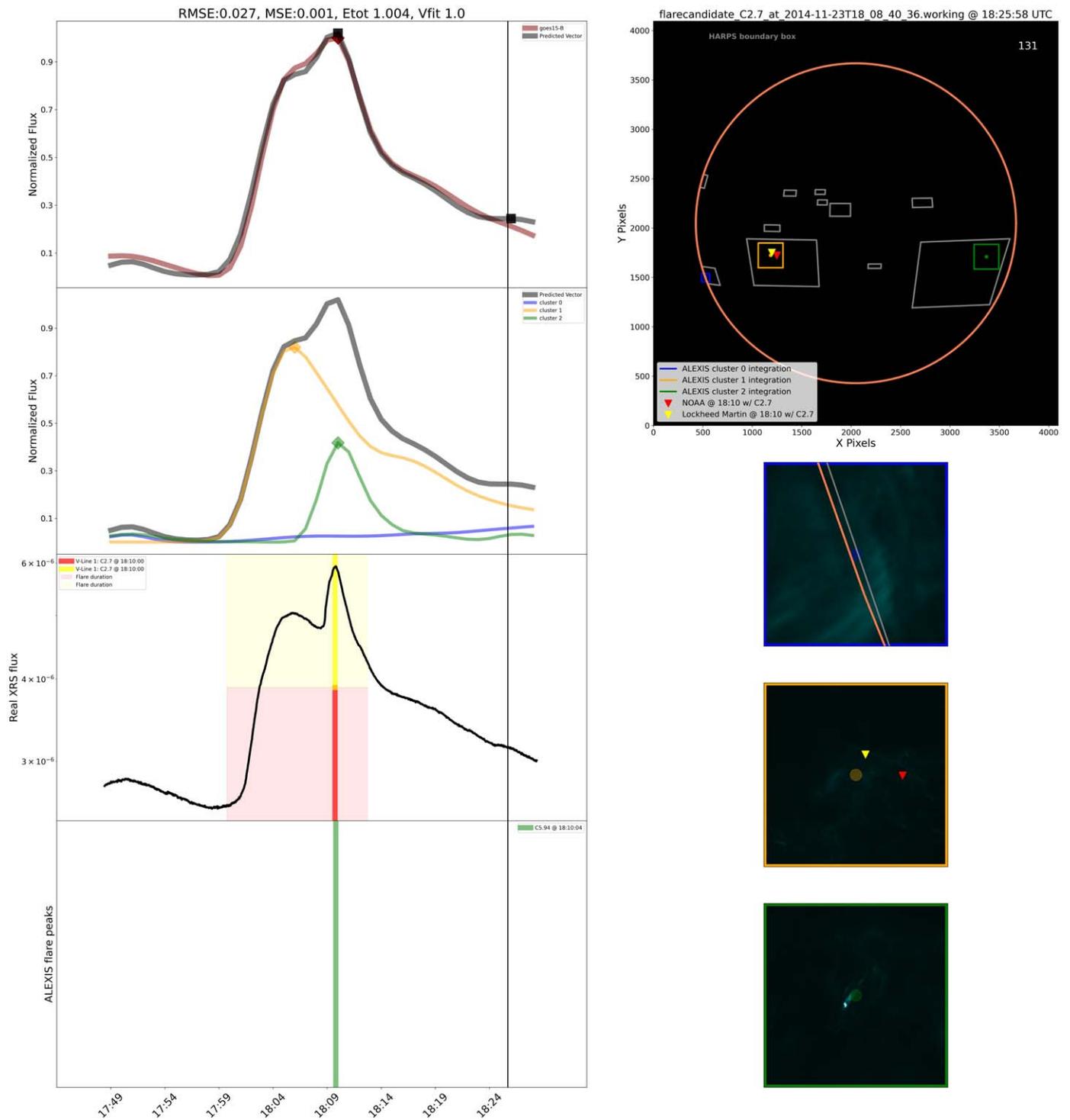

**Figure 17.** Search for C2.7 flare on 2014 November 23 at 18:08:40.36 UTC with GOES15 XRS-B with AIA-131 Å. ALEXIS needs the emission from two regions to recreate the XRS signal; the emission from cluster 2 (green) peaking at 18:10:04 UTC marks the highest data point for the time range under analysis. HARP 4817 and NOAA AR 12209 are responsible for the emission of region 2 (green). Both SolarSoft and the SWPC have reported a C2.7 flare at 18:10:00 UTC occurring in region 1 (orange). They are both accurate and specific in location but they do not report the correct time of the flare. ALEXIS fails to detect the flare occurring in region 1 (orange) at 18:04:00 UTC, which maps to NOAA 12216 and HARP 4851. Note that the magnitude of the only flare detected by ALEXIS is contaminated by the previous emission, another example of how a new labeling scheme independent of the X-ray magnitude of a single pixel is needed to correctly identify the strength of a flare. How does the emission from cluster 1 (orange) influence or increase the probability of magnetic reconnection at the other regions?

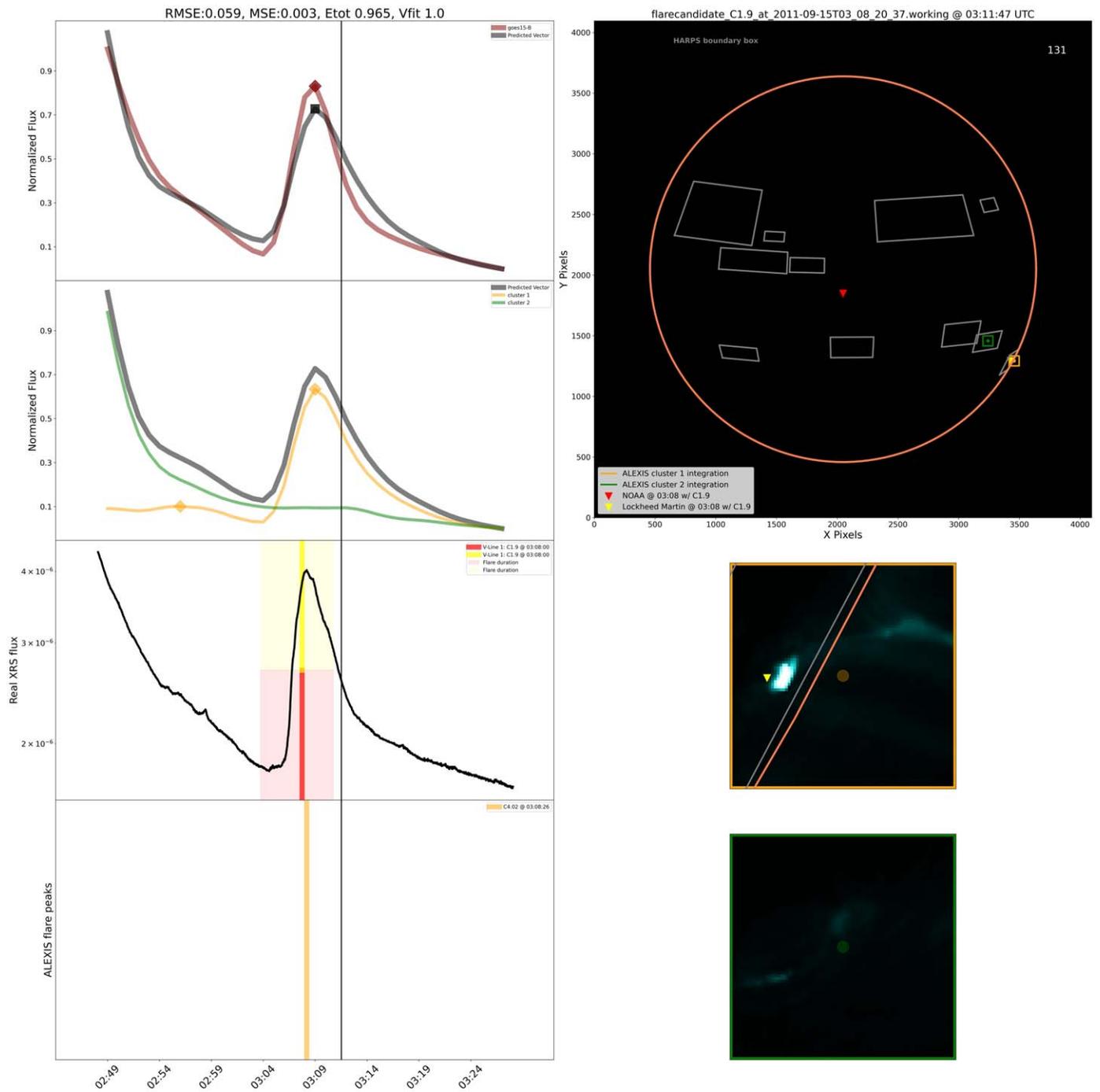

**Figure 18.** Search for C1.9 flare on 2011 September 15 at 03:08:20.37 UTC recreated with GOES15 XRS-B with AIA-131 Å: example where ALEXIS picks up the background emission before the flare. SolarSoft reports a flare that occurred within the ALEXIS integration region for cluster 1 (orange) for a C1.9-class flare on 03:08:00 UTC; SolarSoft is accurate and specific with HARP 843. Note that a NOAA AR was not reported by SolarSoft. The SWPC reports the same flare peak time but has no specific location data and no NOAA AR; for this example, the SWPC is not accurate or specific. The background emission found by ALEXIS can identify where the X-rays are being produced before the actual flare at NOAA AR 11297 associated with HARP 869 and assigned to candidate 2 (green). Can this be a region that can influence global properties of other ARs and increase their probability to undergo energy release?

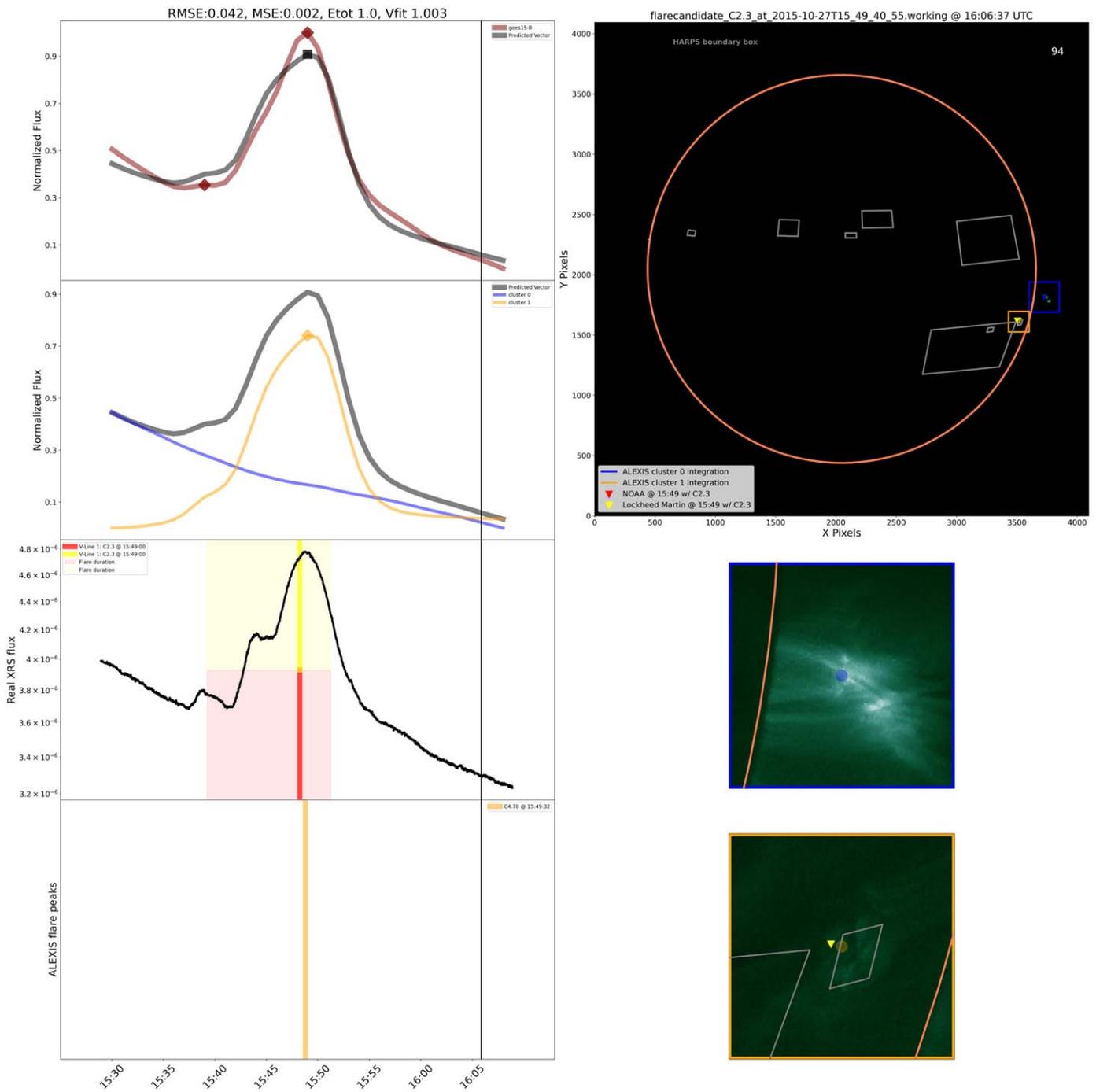

**Figure 19.** Search for C2.3 flare on 2015 October 27 at 15:49:40.55 UTC recreated with GOES15 XRS-B with AIA-94 Å: example where ALEXIS picks up the background emission before and after the flare. SolarSoft and the SWPC report a flare that occurred within the ALEXIS integration region for cluster 1 (orange) for a C2.9-class flare on 15:49:00 UTC at the exact same coordinate. Both teams are specific; if HARP association for these teams were based on exact coordinates, they both would report HARP 6026. Nevertheless, because in practice mapping would be done from the NOAA AR to HARP, their mapping would identify the flare as occurring at NOAA AR 12435 and HARP 6026. We were not able to associate the background emission from region 0 (blue) to any NOAA AR or HARP. An interesting question for events likes these is can we track candidate regions after the flare into a new time range for analysis?

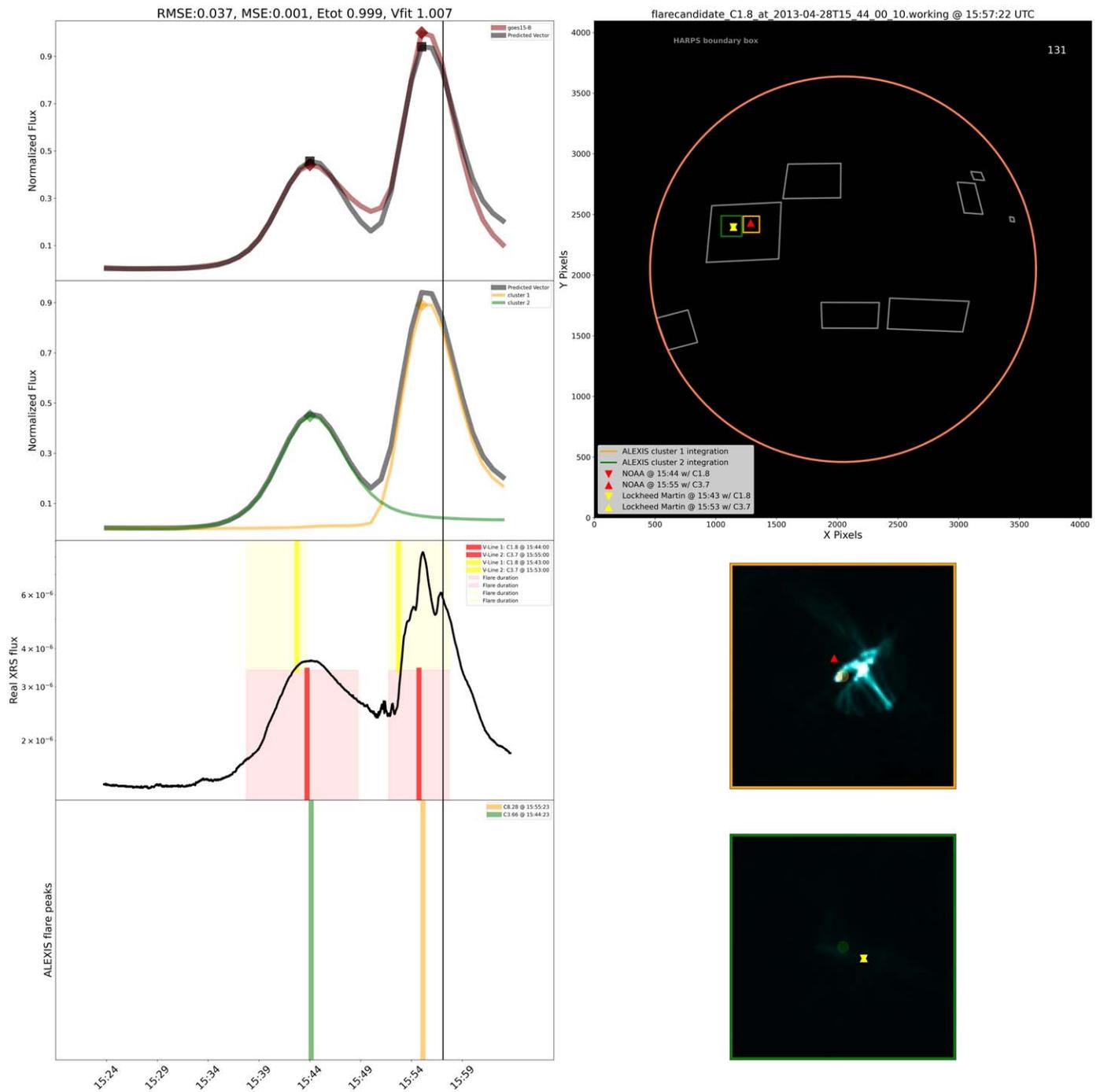

**Figure 20.** Search for C1.8 flare on 2013 April 28 at 15:44:00.10 UTC recreated with GOES15 XRS-B with AIA-131 Å: example that demonstrates the benefits of using the full resolution and cadence of AIA. Two flares within the same AR separated by 10 minutes. SolarSoft identifies the two flares and correctly assigns them to a HARP 2693 and NOAA AR 11731, but SolarSoft's entry at 15:53:00 is not specific. On the other hand, both flares are specific and accurate for the SWPC. Specifying what pixels are representative of flaring regions will help in creating new geometric features.

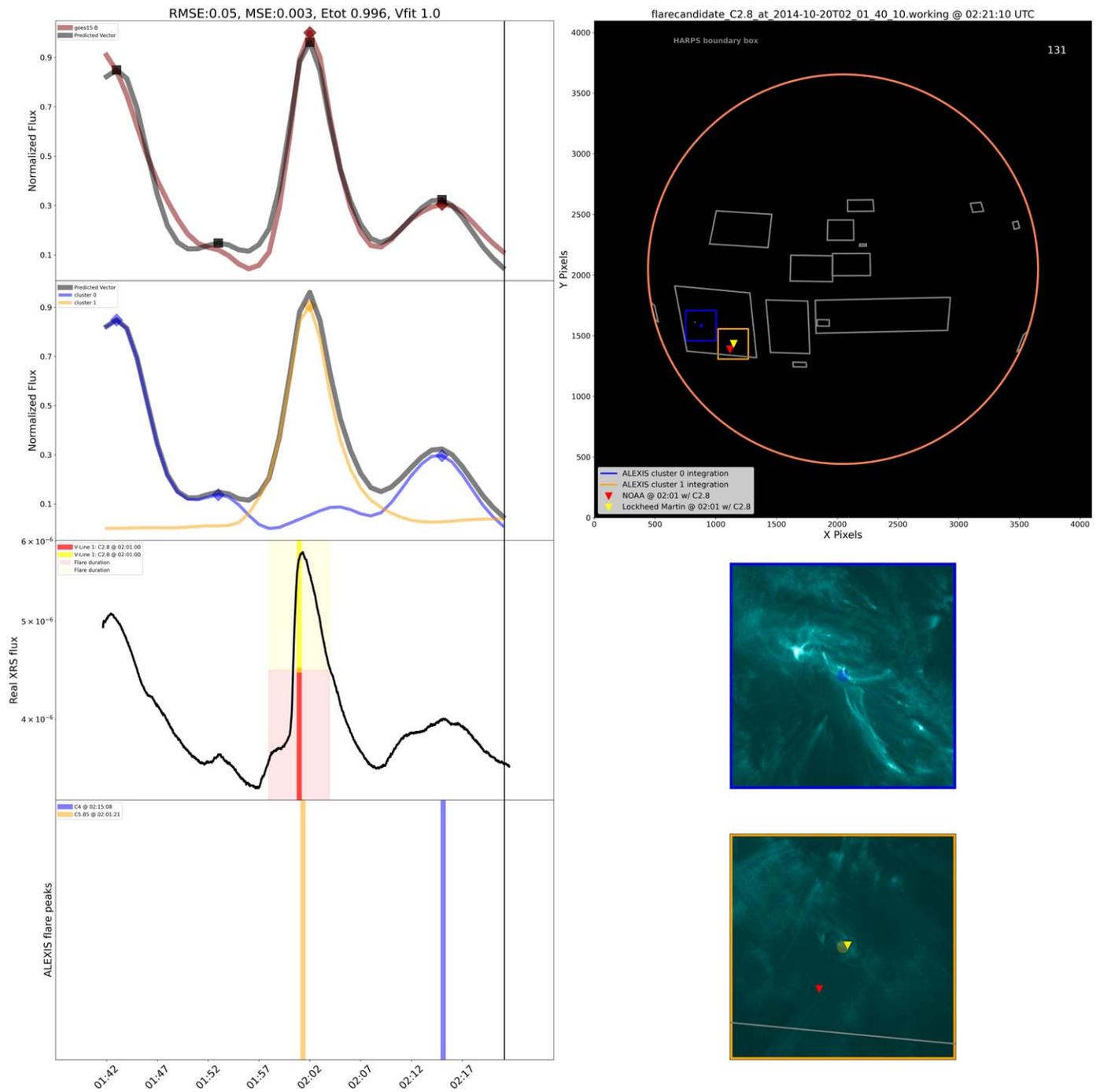

**Figure 21.** Search for C2.8 flare on 2014 October 20 at 02:01:40.10 UTC recreated with GOES15 XRS-B with AIA-131 Å: example that demonstrates the benefits of using the full resolution and cadence of AIA. Two flares within the same AR separated by 13 minutes where one region (blue) can produce the background and also a new flare that was previously uncataloged. SolarSoft and the SWPC identify a flare accurately and specifically at cluster 1 (orange), correctly assigning NOAA AR 12192 and HARP 4698 to the event.

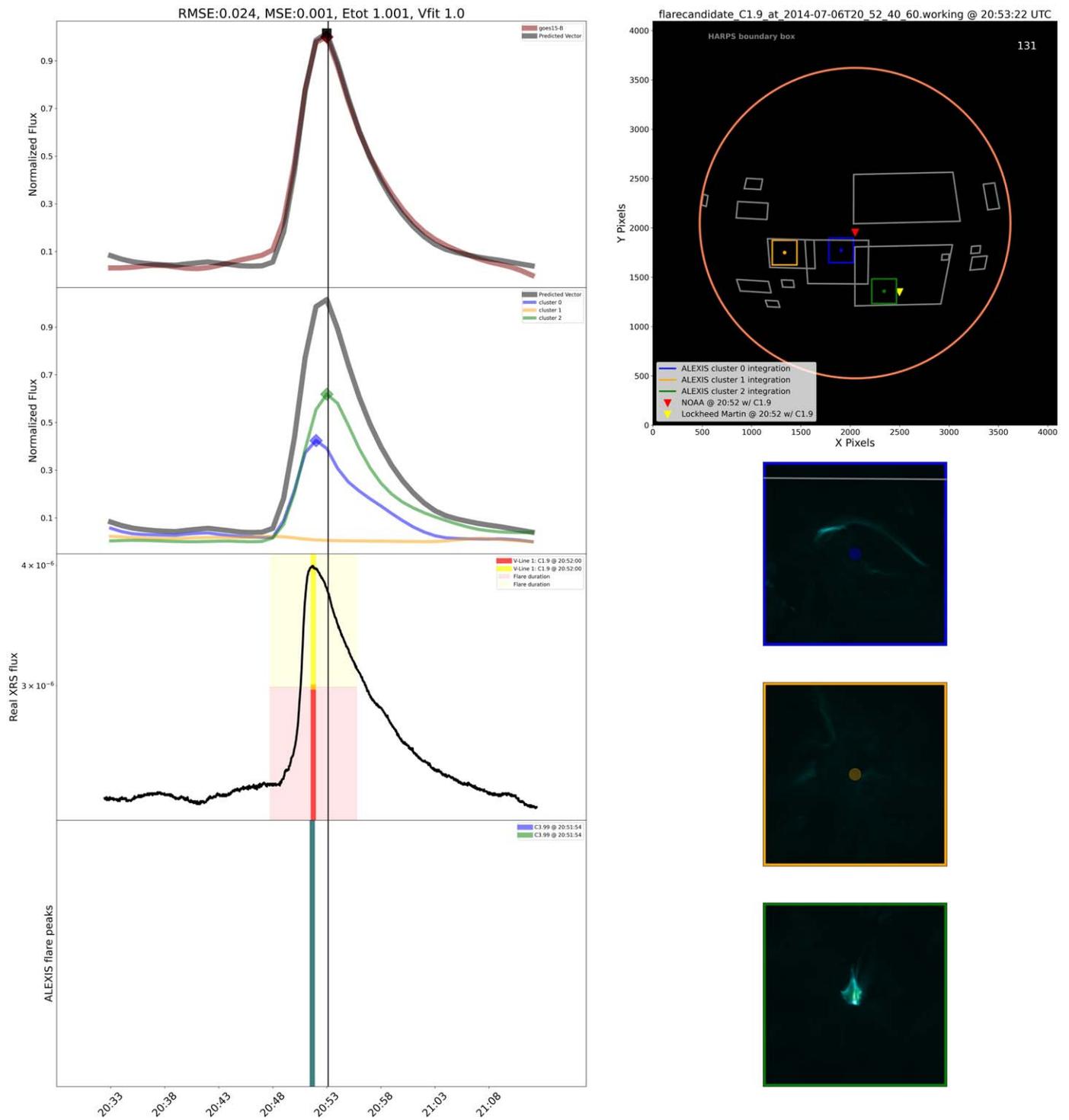

**Figure 22.** Search for flare C1.9 on 2014 July 6 at 20:52:40.60 UTC recreated with GOES15 XRS-B with AIA-131 Å. Sympathetic flares could be evidenced by multiple emission, from different spatial locations independent of each other, which occurs at the same moment. Adding the emission from region 0 (blue, NOAA AR 12108 and HARP 4315) and region 3 (green, NOAA AR 12108 and HARP 4296) will account for the X-rays in this time range. This is an example where the same NOAA AR will map to different HARPs. Both the SWPC and SolarSoft are accurate but not specific reporting the flare as happening at NOAA AR 2107 mapping to HARP 4296 (candidate 3). Can we identify how to assign what the true X-ray class of each of the regions should be constrained by the maximum X-ray emission?

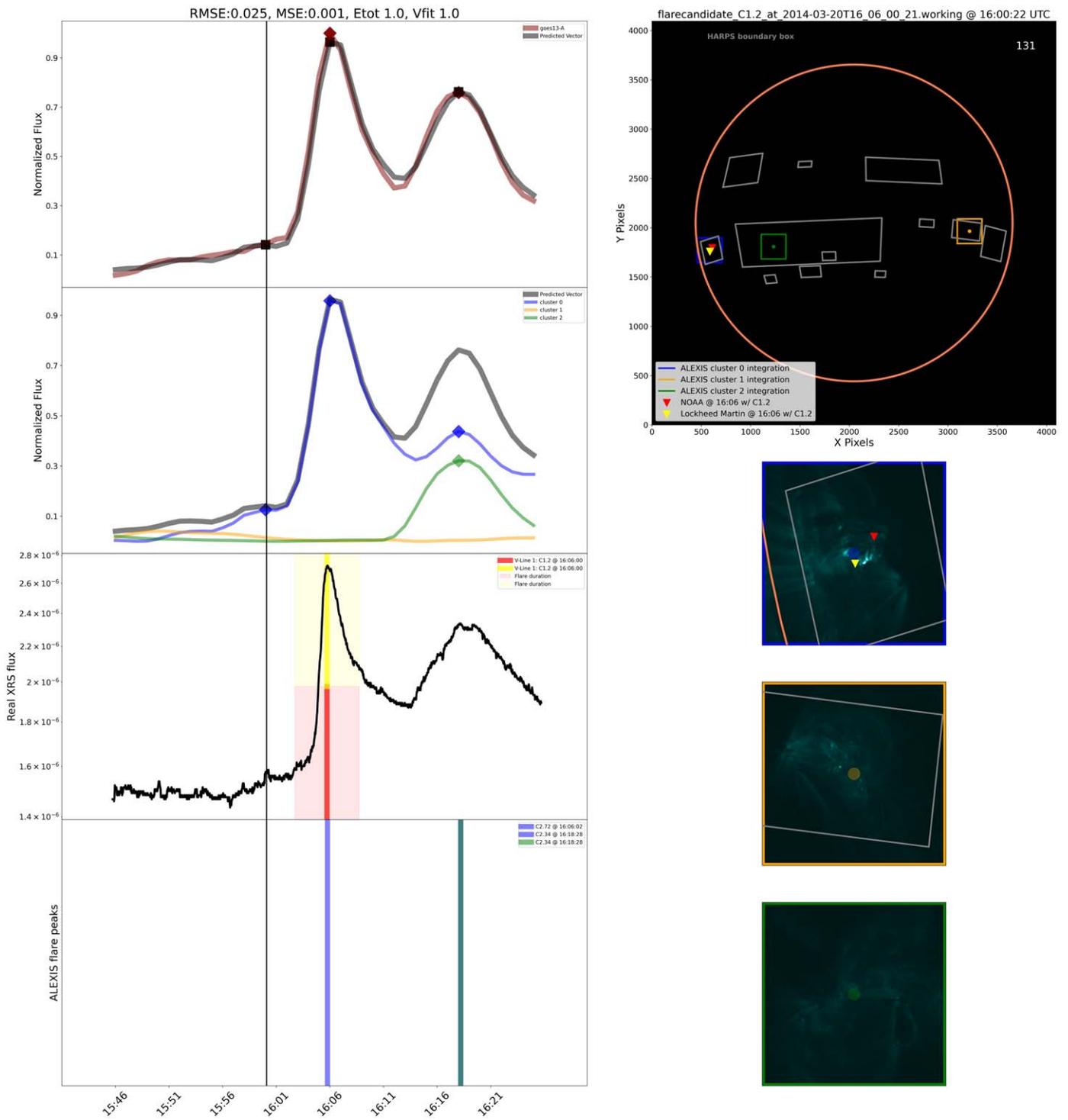

**Figure 23.** Search for flare C1.2 on 2014 March 20 at 16:06:00.21 UTC recreated with GOES13 XRS-A with AIA-131 Å: another example of sympathetic flares. SolarSoft and the SWPC are specific and accurate for a C1.2-class flare that happens at 16:06:00. ALEXIS reports this flare as C2.72. Also, the pipeline is able to identify two more events. These are synchronous flares occurring at 16:18:28 between candidate 0 (blue) and candidate 2 (green).

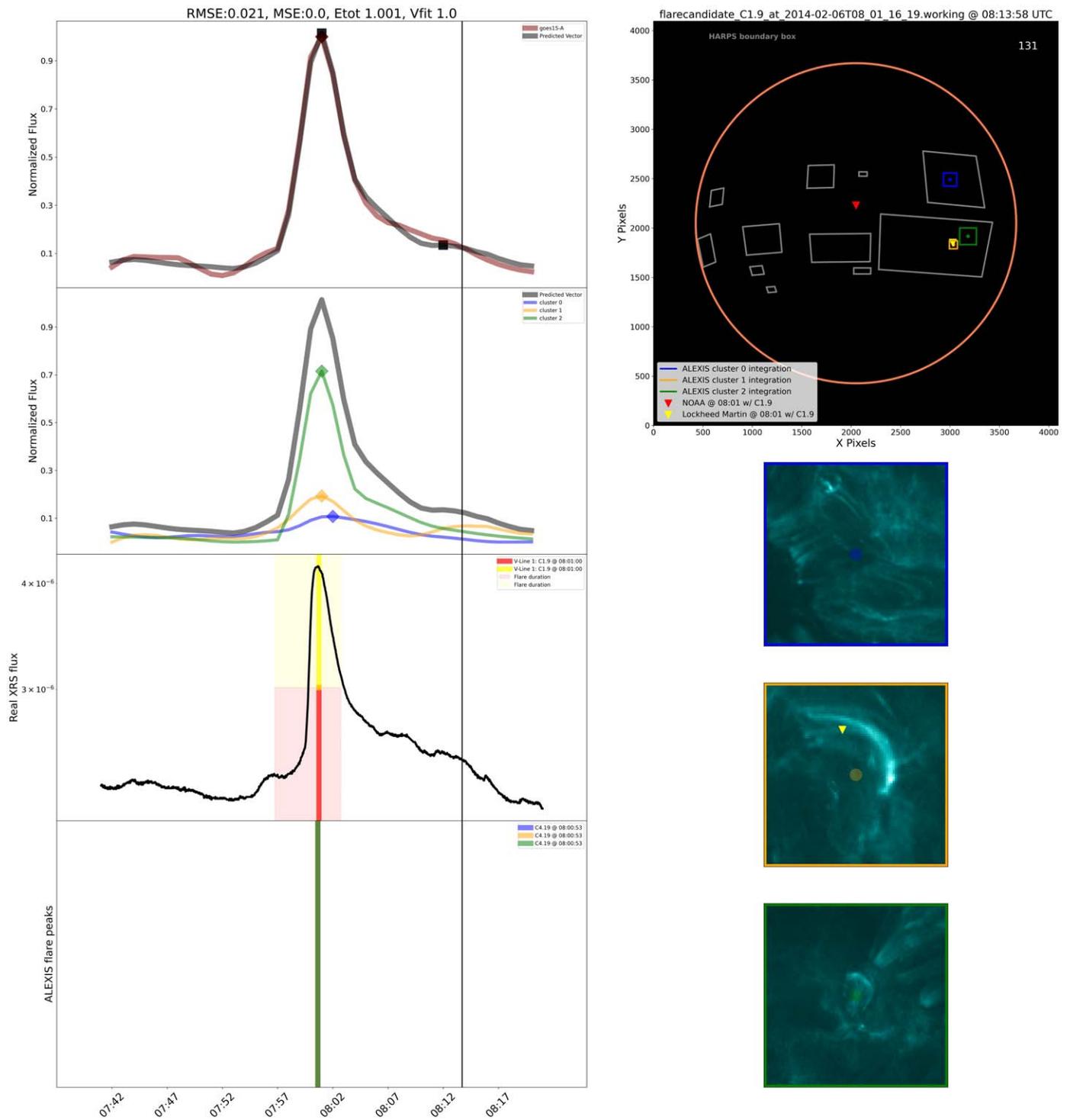

**Figure 24.** Search for flare C1.9 on 2014 February 6 at 08:01:16.19 UTC recreated with GOES15 XRS-A with AIA-131 Å. ALEXIS reports three flares. The SWPC identifies this flare as having happened at NOAA AR 11968, which maps to HARP 3688. This is the same mapping ALEXIS has for region 0; this SWPC entry is accurate but not specific. The metadata for SolarSoft, on the other hand, identifies their flare to have happened in NOAA AR 11967, which maps to HARP 3686; this SolarSoft entry is accurate but not specific for the ALEXIS flare that occurred in region 2 (orange). Candidate 2 (green) has the strongest emission and dominates the flare, followed by candidate 2 (orange) and candidate 0 (blue).

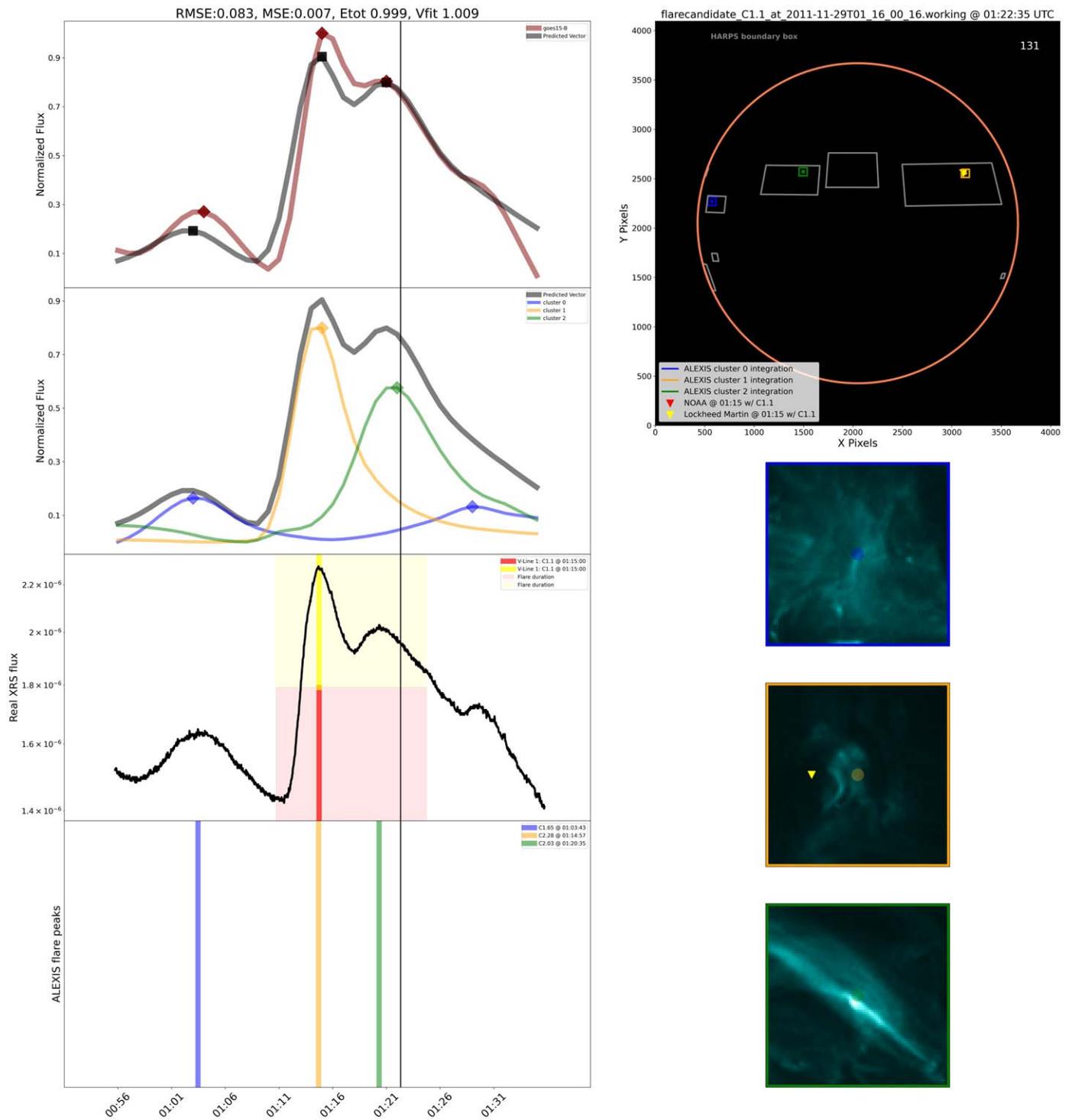

**Figure 25.** Search for flare C1.1 on 2011 November 29 at 01:16:00.16 UTC recreated with GOES15 XRS-B with AIA-131 Å. Three flares and three ARs that span across the face of the solar surface. SolarSoft and the SWPC both have the same metadata for location and timing: an event at candidate 1 (yellow) related to NOAA AR 11356 and to HARP 1093. The decay phase of the flare duration in the canonical metadata fails to catch another flare happening at candidate region 2 (green) related to NOAA AR 11361 and to HARP 1119. Previous to both emissions, candidate 0 (blue) also undergoes energy release related to AR 11362 and HARP 1120.

**ORCID iDs**

Jorge R. Padial Doble  https://orcid.org/0009-0009-3171-7016

Kelly Holley-Bockelmann  https://orcid.org/0000-0003-2227-1322